\documentclass[11pt]{article}
\usepackage{amssymb}
\usepackage{graphics,epsfig,subfigure}
\usepackage{diagbox}
\usepackage{amsthm}
\usepackage{amscd}
\usepackage{amstext}
\usepackage{epstopdf}
\usepackage{multirow}
\newcommand{\Appendix}[1]{
\refstepcounter{section}
\makeatletter
\newcommand{\rmnum}[1]{\romannumeral #1}
\newcommand{\Rmnum}[1]{\expandafter\@slowromancap\romannumeral #1@}
\makeatother
\begin{flushleft}
{\large\bf Appendix \thesection: #1}
\end{flushleft}}

\newcommand{\ba}{\begin{array}}
\newcommand{\ea}{\end{array}}

\let\w=\omega

\def\be{\begin{equation}}
\def\ee{\end{equation}}
\def\ba{\begin{array}}
\def\ea{\end{array}}

\def\dalemb#1#2{{\vbox{\hrule height .#2pt
        \hbox{\vrule width.#2pt height#1pt \kern#1pt
                \vrule width.#2pt}
        \hrule height.#2pt}}}

\def\ocal{{\mathcal{O}}}

\begin{document}
\begin{center}


\vspace{1cm} { \LARGE {\bf Excited states of holographic superconductors from massive gravity}}

\vspace{1.1cm}

Qian Xiang\footnote{xiangq18@.lzu.edu.cn}, Li Zhao\footnote{lizhao@lzu.edu.cn}, Yong-Qiang Wang\footnote{yqwang@lzu.edu.cn, corresponding author}

\vspace{0.7cm}

{\it Institute of Theoretical Physics $\&$ Research Center of Gravitation, Lanzhou University, Lanzhou 730000, People's Republic of China }

\vspace{1.5cm}

\end{center}
\begin{abstract}
\noindent
In this paper, we generalize the study of the model of holographic superconductor in the excited states to the framework of massive gravity. By taking into account the effect of massive graviton, we numerically present a family of solutions of the holographic superconductor in excited states, and find that the critical temperatures can be higher due to the effect of massive graviton comparing with the superconductor in Einstein gravity. We also investigate the condensates and conductivities in the ground state and the excited states with studying various parameters that determine the framework of gravity background.
\end{abstract}

\vspace{5cm}

\pagebreak
\section{Introduction}
In condensed matter physics, Landau-Ginzburg's phenomenological theory and BCS theory suffer from some difficulties to fully explain the mechanism of superconductivity. Due to the strong coupling, electrons can still pair up in some materials with higher critical temperature in which the BCS theory failed to predict. Fortunately, as one of the greatest achievements from the string theory, the AdS/CFT correspondence which establishes the relations between strongly correlated fields on the boundary and weak gravity in an one dimensional higher bulk spacetime, has provided a novel way to study condensed matter physics. In order to model superconductivity, G. T. Horowitz $et~al.$ \cite{Gubser:2008px,Hartnoll:2008vx,Hartnoll:2008kx,Horowitz:2008bn} coupled a complex scalar field to a U(1) gauge field. Considering a spontaneous U(1) symmetry breaking under the critical temperature, they successfully modeled the Cooper pair-like superconductor condensate with the scalar field, which provides us with a new way of understanding the mechanism of high temperature superconductivity. Afterwards, the holographic superconductor condensates were realized in various ways. By replacing the complex field with a symmetric and traceless second-rank tensor, the holographic d-wave model was constructed in \cite{Chen:2010mk,Benini:2010pr,Kim:2013oba}. The holographic p-wave superconductor condensate can be modeled by a two-form field \cite{Aprile:2010ge} or a complex, massive vector field with U(1) charge \cite{Cai:2013pda,Cai:2013aca}. In addition, the holographic p-wave superconductor can also be constructed by the SU(2) Yang-Mills field coupling to  gravity \cite{Gubser:2008wv}.
For reviews of holographic superconductor, see Refs. \cite{Hartnoll:2009sz,Herzog:2009xv,Horowitz:2010gk,Cai:2015cya}.

Recently, inspired by the seminal paper of Gubser $et~al.$ \cite{Gubser:2008px}, Yong-Qiang Wang $et~al.$ generalized the the studies of holographic superconductor from ground state to excited states. In \cite{Wang:2019caf}, the author found a family of solutions of excited states for the s-wave holographic superconductor in the probe limit, where the scalar field possesses $n$ nodes along the radial coordinate corresponding to $n$-th excited state. Moreover, in \cite{Wang:2019vaq}, considering the holographic superconductor with the full backreaction, the author also numerically studied the condensate and conductivity of their ground state and excited states. It was found that the high excited states of $\langle \ocal_1 \rangle$ condensates converge to about $4.4$ as the temperature approaching zero. Besides, it was also found that the condensation values of $ \ocal_2 $ operator in the excited states are larger than the ground state. Other studies on the excited state of holographic superconductor can be found in \cite{Qiao:2020fiv,Li:2020omw}, where in \cite{Qiao:2020fiv}, the author proposed a analytic technique based on variational method for the Sturm-Liouville eigenvalue problem, to investigate the excited states of the holographic superconductor in the probe limit. The non-equilibrium process of the holographic s-wave superconductor with the excited states was investigated in \cite{Li:2020omw}.

In the above discussion, the Cooper pair-like superconductor condensate was modeled by matter fields coupling to Einstein gravity. As a natural curiosity, it is always interesting to ask how will such a system behaves, if the graviton is massive. However this is not an easy task, since the diffeomorphism invariance in GR has already restrict the graviton to be the massless spin-2 boson \cite{17}. Simple generalization would usually make the massive gravity unstable because of the problem of Boulware-Deser ghost \cite{BDG}. To avoid this, C. de Rham and G. Gabadadze (dRGT) introduced polynomial terms in general action successfully building a ghost-free nonlinear massive gravity theory \cite{deRham:2010ik,Hinterbichler:2011tt,deRham:2010kj}. Such novel achievement arouse great concerns in the realm of holographic superconductivity. The holographic DC and Hall conductivity was studied in \cite{Zhou:2015dha} in massive Einstein-Maxwell-Dilaton gravity providing a holographic method to explain strange metals. The nonequilibrium process of holographic s-wave superconductor was investigated in the dRGT massive gravity in \cite{Li:2019qkt}. In \cite{Nam:2020bfq}, Cao H. Nam investigated the effect of massive gravity on the p-wave holographic superconductor and found that the critical temperature and condensate dependent crucially on the sign of the massive gravity couplings. In addition, attempting to avoid complicated numerical calculation, a framework for translational symmetry breaking and momentum dissipation was set up in \cite{Vegh:2013sk}. In \cite{Hu:2015dnl}, Ya-Peng Hu studied a holographic Josephson junction of its critical temperature, tunneling current and coherence length, and find that due to the graviton mass, the transition from normal state to superconducting state would be harder. The model of holographic superconductor with backreaction in massive gravity can be found at \cite{Zeng:2014uoa}. Besides, a recent experiment conducted by LIGO on the detection of gravitational waves produced by a binary black hole merger provided an upper bound of graviton mass that $m_g \leq 1.2\times10^{-22}$ \cite{LIGO}.

However, until now, the investigation on holographic superconductor of its condensate and conductivity in the ground state and the excited states has not yet been studied in the framework of massive gravity. Therefore, in this paper, we will numerically investigate the holographic superconductor of its condensate and conductivity in the ground state and the excited states with considering the effect of massive graviton.

The paper is arranged as follows: in section \ref{review} we give a brief review of the dRGT massive gravity. The construction of holographic superconductor in $(3+1)$-dimensional AdS spacetime with massive graviton can be found at section \ref{setup}. The numerical results and properties of scalar condensate and optical conductivity are in section \ref{results}. The conclusion and discussion are in the last section.

\section{Review of dRGT Massive Gravity}\label{review}
In order to construct a holographic superconductor in the framework of massive gravity we consider the following action in a $4$-dimensional ghost-free dRGT massive gravity, which is the Hilbert-Einstein action including the nonlinear polynomial terms \cite{deRham:2010ik}
\begin{equation}
\label{eq1}
S =\frac{1}{2\kappa^2}\int d^{4}x \sqrt{-g} [ R +\frac{6}{\ell^2} +\lambda^2 \sum^4_i c_i {\cal U}_i (g,f)],
\end{equation}
where $f$ is the reference metric, the bulk metric is denoted by $g$ and the nonzero graviton mass is represented by $\lambda$. $c_i$ are constants, $\ell$ is the length scale of the AdS spacetime, and $R$ is the usual Ricci tensor. The ${\cal U}_i$ are nonlinear interaction terms of the eigenvalues of the $4\times4$ matrix ${\cal K}^{\mu}_{\ \nu} \equiv \sqrt {g^{\mu\alpha}f_{\alpha\nu}}$:
\begin{eqnarray}
\label{eq2}
&& {\cal U}_1= [{\cal K}], \nonumber \\
&& {\cal U}_2=  [{\cal K}]^2 -[{\cal K}^2], \nonumber \\
&& {\cal U}_3= [{\cal K}]^3 - 3[{\cal K}][{\cal K}^2]+ 2[{\cal K}^3], \nonumber \\
&& {\cal U}_4= [{\cal K}]^4- 6[{\cal K}^2][{\cal K}]^2 + 8[{\cal K}^3][{\cal K}]+3[{\cal K}^2]^2 -6[{\cal K}^4],
\end{eqnarray}
where the ${\cal K}$ stands for $(\sqrt{A})^{\mu}_{\ \nu}(\sqrt{A})^{\nu}_{\ \lambda}=A^{\mu}_{\ \lambda}$ and $[{\cal K}]=K^{\mu}_{\ \mu}$. By varying the above action (\ref{eq1}) with respect to the metric $g_{\mu\nu}$,
the equations of motion (EoM) turns out to be
\begin{eqnarray}
R_{\mu\nu}-\frac{1}{2}Rg_{\mu\nu}-\frac{3}{\ell^2} g_{\mu\nu}+\lambda^2 \chi_{\mu\nu}=8\pi G T_{\mu \nu },~~
\end{eqnarray}
where
\begin{eqnarray}
 \chi_{\mu\nu}=-\frac{c_1}{2}({\cal U}_1\,g_{\mu\nu}-{\cal K}_{\mu\nu})-\frac{c_2}{2}({\cal U}_2\,g_{\mu\nu}-2\,{\cal U}_1{\cal K}_{\mu\nu}+2{\cal K}^2_{\mu\nu})
-\frac{c_3}{2}({\cal U}_3\,g_{\mu\nu}-3\,{\cal U}_2{\cal K}_{\mu\nu}\nonumber \\
+6\,{\cal U}_1{\cal K}^2_{\mu\nu}-6{\cal K}^3_{\mu\nu})
-\frac{c_4}{2}({\cal U}_4\,g_{\mu\nu}-4\,{\cal U}_3{\cal K}_{\mu\nu}+12\,{\cal U}_2{\cal K}^2_{\mu\nu}-24\,{\cal U}_1{\cal K}^3_{\mu\nu}+24{\cal K}^4_{\mu\nu}).
\end{eqnarray}

Since we study the holographic superconductor in $4$ dimensional background, we choose the reference metric as:
\begin{equation}
\label{reference}
f_{\mu\nu} = {\rm diag}(0,\; 0,\; c_0^2 h_{ij} ),
\end{equation}
where $c_0$ is a positive constant.
A general black hole solution can be found~\cite{Cai:2014znn}
\begin{eqnarray}\label{metric}
ds^2=-r^{2}f(r)dt^{2}+\frac{dr^{2}}{r^{2}f(r)}+r^{2}h_{ij}dx^{i}dx^{j},
\end{eqnarray}
with
\begin{eqnarray}\label{metric}
\label{black hole solution} f(r)&=&\frac{k}{r^2}+\frac{1}{\ell^2}-\frac{m_0}{r^3}+\frac{q^2}{4r^4}+\frac{c_1\lambda^2}{2r}+
\frac{c_2\lambda^2}{r^2}.
\end{eqnarray}
Here, $m_0$ is the black hole mass and the charge of it is denoted by $q$. The $c_1$ and $c_2$ can be regarded as the coupling coefficients with graviton mass $\lambda$. We choose $k=0$ in this paper, and hence, the line element $h_{ij}dx^{i}dx^{j}$ is related to a two dimensional flat space.

The Hawking temperature of the black hole solution is given by
\begin{equation}
T_{BH}=\frac{\left(r^2f(r)\right)'}{4\pi}\bigg|_{r=r_h}=\frac{1}{4\pi r_h}\left(k+\frac{3r_h^2}{\ell^2}-\frac{q^2}{4r_h^2}+c_1\lambda^2r_h +c_2\lambda^2\right),
\end{equation}
which can be regarded as the temperature of the holographic superconductor. The $r_h$ is the radius of the black hole, which is numerically set to $r_h=1$. Without loss of generality, we also set $q=0$ and the AdS length scale $\ell=1$ in the following numerical calculations.
\section{Holographic Setup}\label{setup}
In order to model the scalar condensate at low temperature, we consider the following Maxwell and complex scalar field action in $\textrm{AdS}_4$ as
\begin{equation}\label{action2}
S=\int d^{4} x\sqrt{-g}\left(-\frac{1}{4}F_{\mu\nu}F^{\mu\nu}-|\nabla\psi-iA\psi|^2-\alpha^2|\psi|^2\right),
\end{equation}
in which $A$ is the $U(1)$ gauge field with the corresponding field strength $F_{\mu\nu}=\partial_\mu A_\nu-\partial_\nu A_\mu$. $\alpha$ is the mass of the complex scalar field $\psi$. Note that, we have rescaled the above action with the charge of the scalar field and therefore the backreaction of the matter field on the metric can be negligible if the charge is large enough.

From the above action, the EoMs can be obtained as
\begin{eqnarray}
&(\nabla_\mu-iA_\mu)(\nabla^\mu-iA^\mu)\psi-\alpha^2\psi=0, \\
&\nabla_\nu F^{\nu\mu}=i(\psi^*(\nabla^\mu-iA^\mu)\psi-\psi(\nabla^\mu+iA^\mu)\psi^*).
\end{eqnarray}
On the two dimensional flat space, an isotropic ansatz for the U(1) gauge field and the scalar field can be written as
\begin{equation}\label{ansatz}
  A=\phi(r) dt, \;\;\;\;\; \psi=\psi(r).
\end{equation}
With the above ansatz, the equations of motion for the scalar field $\psi(r)$ and electrical scalar potential $\phi(r)$ in the
background of the Schwarzschild-AdS black hole are
\begin{eqnarray}
\psi'' + \left(\frac{f'}{f} + \frac{2}{r}\right) \psi' +\frac{\phi^2}{f^2}\psi - \frac{\alpha^2}{\ell^2 f} \psi &=& 0 \,,\label{eom1}\\
\phi'' + \frac{2}{r} \phi' - \frac{2\psi^2}{f} \phi &=& 0.\label{eom2}
\end{eqnarray}
Near the infinite boundary $r\to\infty$, the asymptotic behaviors of the functions $\psi(r)$ and  $\phi(r)$ have the following forms
\begin{eqnarray}\label{asympphi}
\psi &=& \frac{\psi^{(1)}}{r^{\Delta_-}} + \frac{\psi^{(2)}}{r^{\Delta_+}} + \cdots \,,\\
\phi &=& \mu - \frac{\rho}{r} + \cdots \,,
\end{eqnarray}
with
\begin{equation}\label{delta1}
\Delta_\pm=\frac{3\pm\sqrt{9+4 \alpha^2}}{2}.
\end{equation}
According to the AdS/CFT dictionary, $\psi^{(i)} (i=1,2) $ are the corresponding expectation value of $\langle \ocal_i\rangle$ condensate of the dual scalar operators $\mathcal{O}_{i}$.  And $\mu$, $\rho$ denote chemical potential and charge density in the boundary field theory, respectively. Note that the mass of the complex field $\alpha$ will have to satisfy the Breitenlohner-Freedman(BF) bound of $\alpha^2\geq -9/4$ \cite{BF}. For simplicity, we will set $\alpha^2=-2$ in this paper.
\section{Numerical Results}\label{results}
In condensed matter physics, when temperature $T$ drops below a critical temperature $T_c$, electrical resistivity will suddenly drops to zero leading to superconductivity. In the holographic approach, due to the spontaneously broken of U(1) gauge symmetry, the scalar condensate will turn on when the chemical potential $\mu$ is larger than $\mu_c$. Besides, because the holographic superconductor is a strongly coupled system, the quasi-particles will interact with each other and form a bound state, which eventually leads to the excited states of the scalar condensate. In this section, we will numerically study the holographic superconductor in the framework of massive gravity of its condensate and conductivity in ground and excited states.

Firstly, we conduct a simple coordinate transformation, which reads:
\begin{equation}\label{trans}
 r=r_h/z.
\end{equation}
We will set the spatial infinity $z_ \infty=0$ and the black hole radius $z_h=1$ for simplicity. After imposing the above transformation, we conduct all of our numerical calculations based on spectral method. In the integration region $0 \leq z \leq 1$, our choice of typical grids have the sizes from 50 to 300.

In order to study the scalar condensate of its ground and excited states for both $\ocal_1 $ and $ \ocal_2 $ operators, the coupled equations (\ref{eom1}) and (\ref{eom2}) will be put into an iterative process by means of Newton-Raphson method. We will give a good initial guess that leads to a solution for ground or $n$-excited state for each time. For the ground state of $\langle \ocal_i \rangle$ condensate, the initial guess for $\psi(z)$ is the profile that have no node along the $z$ coordinate. For the $n$-th excited state, we choose the initial guess that allows $\psi(z)$ to have $n$ nodes on the $z$ axis to reach the excited state. The estimated relative error for our numerical results is below $10^{-5}$.

Recall that the effect of massive gravity to the holographic superconductor is involved in the black hole solution (\ref{black hole solution}) performing by the coupling coefficients $c_1$, $c_2$ and the graviton mass $\lambda$. Numerically changing their values, will enable us to study the behaviours of such holographic system of its scalar condensate and its conductivity under different configurations of massive gravity framework. However, the contribution of massive gravity will disappear and the situation will go back to \cite{Hartnoll:2008vx,Wang:2019caf} by setting these parameters to be $c_1=0$, $c_2=0$ or simply, $\lambda=0$. Therefore, in the study of the effect of $c_1$ (or $c_2$) we change $c_1$ (or $c_2$) only with fixing $c_2=0$ (or $c_1=0$) and $\lambda\neq 0$. As for studying graviton mass, we change $\lambda$ only while maintaining $c_1=1,~c_2=-0.5$.

\subsection{Critical Chemical Potential}
In Tab. \ref{table c1}, Tabs. \ref{table c2} and \ref{table m}, we present our results of critical chemical potential $\mu_{c}^{n}$ from ground state to fifth excited state. In Tabs. \ref{table c1} and \ref{table c2}, we investigate the coupling coefficients $c_1$ and $c_2$ changing from $-1$ to $3$ solely. In Tab. \ref{table m}, we investigate the graviton mass changing from $0.1$ to $0.5$ solely.

From the three tables, we find that the $\mu_{c}$ increases with the growth of $n$-th excited state. Moreover, as we enlarge the coupling coefficients and graviton mass, the critical chemical potentials of each state also increase for both kinds of condensates.
\begin{table}[!h]
\begin{center}
\begin{tabular}{c |c c c| c c c c c c c}
\hline
$\langle \ocal_i \rangle$~&~$c_1$~&~$c_2$~&~$\lambda$~&~$\mu_{c}^{0}$~&~$\mu_{c}^{1}$~&~$\mu_{c}^{2}$~&~$\mu_{c}^{3}$~&~$\mu_{c}^{4}$~&~$\mu_{c}^{5}$\\
\hline
\multirow{3}*{$\langle \ocal_1 \rangle$}&-1&0&0.2
& 0.668 & 6.383 & 11.585 & 16.759 & 21.927 & 27.092 \\
\cline{2-10}
 &1&0&0.2& 1.429 & 6.603 & 11.816 & 17.037 & 22.262 & 27.488 \\
\cline{2-10}
 &3&0&0.2& 1.884 & 6.820 & 12.045 & 17.312 & 22.593 & 29.710 \\
\hline
\multirow{3}*{$\langle \ocal_2 \rangle$}
&-1&0&0.2& 4.014 & 9.113 & 14.251 & 19.400 & 24.554 & 29.710 \\
\cline{2-10}
 &1&0&0.2& 4.114 & 9.263 & 14.462 & 19.676 & 24.897 & 30.120 \\
\cline{2-10}
 &3&0&0.2& 4.213 & 9.411 & 14.672 & 19.950 & 25.237 & 30.527 \\
\hline
\end{tabular}\caption{The table refers to critical chemical potential $\mu_c^{n}$ of holographic superconductor from ground state to fifth excited state for the coupling coefficient $c_1$, where only $c_1$ changes its values and $c_2$ and $\lambda$ remain fixed.}\label{table c1}
\end{center}
\end{table}

\begin{table}[!h]
\begin{center}
\begin{tabular}{c |c c c| c c c c c c c}
\hline
$\langle \ocal_i \rangle$~&~$c_1$~&~$c_2$~&~$\lambda$~&~$\mu_{c}^{0}$~&~$\mu_{c}^{1}$~&~$\mu_{c}^{2}$~&~$\mu_{c}^{3}$~&~$\mu_{c}^{4}$~&~$\mu_{c}^{5}$\\
\hline
\multirow{3}*{$\langle \ocal_1 \rangle$}&0&-1&0.2
& 1.098 & 6.456 & 11.642 & 16.817 & 21.989 & 27.161\\
\cline{2-10}&0&1&0.2
& 1.142 & 6.530 & 11.759 & 16.980 & 22.199 & 27.418\\
\cline{2-10}
 &0&3&0.2& 1.184 & 6.603 & 11.875 & 17.141 & 22.406 & 27.672\\
\hline
\multirow{3}*{$\langle \ocal_2 \rangle$}
&0&-1&0.2& 4.037 & 9.140 & 14.287 & 19.445 & 24.681 & 29.775\\
\cline{2-10}
 &0&1&0.2& 4.091 & 9.236 & 14.527 & 19.632 & 24.842 & 30.055\\
\cline{2-10}
 &0&3&0.2& 4.144 & 9.329 & 14.565 & 19.815 & 25.072 & 30.332\\
\hline
\end{tabular}\caption{The table refers to critical chemical potential $\mu_c^{n}$ of holographic superconductor from ground state to fifth excited state for the coupling coefficient $c_2$, where only $c_2$ changes its values and $c_1$ and $\lambda$ remain fixed.}\label{table c2}
\end{center}
\end{table}

\begin{table}[t]
\begin{center}
\begin{tabular}{c |c c c| c c c c c c c}
\hline
$\langle \ocal_i \rangle$~&~$c_1$~&~$c_2$~&~$\lambda$~&~$\mu_{c}^{0}$~&~$\mu_{c}^{1}$~&~$\mu_{c}^{2}$~&~$\mu_{c}^{3}$~&~$\mu_{c}^{4}$~&~$\mu_{c}^{5}$\\
\hline
\multirow{3}*{$\langle \ocal_1 \rangle$}&1&-0.5&0.1
& 1.203 & 6.516 & 11.722 & 16.923 & 22.123 & 27.324\\
\cline{2-10}
 &1&-0.5&0.3& 1.712 & 6.698 & 11.894 & 17.119 & 22.353 & 27.591\\
\cline{2-10}
 &1&-0.5&0.5& 2.371 & 7.053 & 12.236 & 17.508 & 22.809 & 28.123\\
\hline
\multirow{3}*{$\langle \ocal_2 \rangle$}
&1&-0.5&0.1& 4.073 & 9.201 & 14.375 & 19.562 & 24.754 & 29.950\\
\cline{2-10}
 &1&-0.5&0.3& 4.146 & 9.303 & 14.516 & 19.744 & 24.980 & 30.219\\
\cline{2-10}
 &1&-0.5&0.5& 4.291 & 9.506 & 14.796 & 20.108 & 25.429 & 30.756\\
\hline
\end{tabular}\caption{The table refers to critical chemical potential $\mu_c^{n}$ of holographic superconductor from ground state to fifth excited state for the graviton mass $\lambda$, where only $m$ changes its values and $c_1$ and $c_2$ remain fixed.}\label{table m}
\end{center}
\end{table}

Besides, we also find that the difference of $\mu_c$ between consecutive states in the framework of massive gravity is also about $5$ for both $\langle \ocal_1 \rangle$ and $\langle \ocal_2 \rangle$, which is accord with the a semi-analytical results in \cite{Qiao:2020fiv}. Therefore, we fit the the relations between state $n$ and $\mu_c$, the results are as follows.

The fittings of coupling coefficient $c_1$, listed in Tab. \ref{table c1}
\begin{equation}
\mu_c\approx\left\{ \begin{array}{ll}
5.255\, n+ 0.932\,, & \quad\ c_1=-1, \, c_2=0,\, \lambda=0.2\,,  \\
5.260\, n+ 1.198\,, & \quad\ c_1=\,\,\,1, \, c_2=0,\, \lambda=0.2\,, \,\,\,\,   \qquad\textrm{for} \;\;  \ocal_1,\\
5.216\, n+ 1.715\,, & \quad\ c_1=\,\,\,3, \, c_2=0,\, \lambda=0.2\,.
\end{array}\label{c1o1} \right.
\end{equation}
\begin{equation}
\mu_c\approx\left\{ \begin{array}{ll}
5.142\, n+ 3.986\,,  & \quad\ c_1=-1,\, c_2=0,\, \lambda=0.2,\\
5.204\, n+ 4.078\,,  & \quad\ c_1=\,\,\,1,\, c_2=0,\, \lambda=0.2,\,   \qquad\textrm{for} \;\;  \ocal_2,\\
5.266\, n+ 4.169\,,  & \quad\ c_1=\,\,\,3,\, c_2=0,\, \lambda=0.2.
 \end{array} \right.
\end{equation}

and the fittings of coupling coefficient $c_2$, listed in Tab. \ref{table c2}
\begin{equation}
\mu_c\approx\left\{ \begin{array}{ll}
5.203\, n+ 1.188\,,  & \quad\ c_1=-1,\, c_2=0,\, \lambda=0.2,\\
5.246\, n+ 1.223\,,  & \quad\ c_1=\,\,\,1,\, c_2=0,\, \lambda=0.2,\,   \qquad\textrm{for} \;\;  \ocal_1,\\
5.289\, n+ 1.258\,,  & \quad\ c_1=\,\,\,3,\, c_2=0,\, \lambda=0.2.
 \end{array} \right.
\end{equation}

\begin{equation}
\mu_c\approx\left\{ \begin{array}{ll}
5.156\, n+ 4.003\,,  & \quad\ c_1=-1,\, c_2=0,\, \lambda=0.2,\,   \\
5.196\, n+ 4.058\,,  & \quad\ c_1=\,\,\,1,\, c_2=0,\, \lambda=0.2,\,\,   \qquad\textrm{for} \;\;  \ocal_2,\\
5.240\, n+ 4.108\,,  & \quad\ c_1=\,\,\,3,\, c_2=0,\, \lambda=0.2.
 \end{array} \right.
\end{equation}

Similarly, the fittings of graviton mass $\lambda$, listed in Tab. \ref{table m}
\begin{equation}\label{exception m}
\mu_c\approx\left\{ \begin{array}{ll}
5.218\, n+ 1.257\,,  & \quad\ c_1=-1,\, c_2=0,\, \lambda=0.2,\,   \\
5.188\, n+ 1.591\,,  & \quad\ c_1=\,\,\,1,\, c_2=0,\, \lambda=0.2,\,\,   \qquad\textrm{for} \;\;  \ocal_1,\\
5.180\, n+ 2.067\,,  & \quad\ c_1=\,\,\,3,\, c_2=0,\, \lambda=0.2.
 \end{array} \right.
\end{equation}

\begin{equation}
\mu_c\approx\left\{ \begin{array}{ll}
5.178\, n+ 4.041\,,  & \quad\ c_1=-1,\, c_2=0,\, \lambda=0.2,\,   \\
5.218\, n+ 4.107\,,  & \quad\ c_1=\,\,\,1,\, c_2=0,\, \lambda=0.2,\,\,   \qquad\textrm{for} \;\;  \ocal_2,\\
5.297\, n+ 4.238\,,  & \quad\ c_1=\,\,\,3,\, c_2=0,\, \lambda=0.2.
 \end{array} \right.
\end{equation}
\subsection{Condensate}
In this subsection, we study the condensates of our holographic superconductor for both $\ocal_1$ and $\ocal_2$ operators in the framework of massive gravity. According to AdS/CFT dictionary, the expectation value of $\langle \ocal_i \rangle$ dual to $\psi^{(i)}$ is
\begin{equation}\label{oi}
\langle \ocal_i \rangle = \sqrt{2}\psi(z)^{(i)}, i=1,2.
\end{equation}
As usual, we study the effects of massive graviton on scalar condensates by numerically solving the Eq. (\ref{eom1}), Eqs. (\ref{eom2}) and (\ref{oi}) with changing parameters $c_1,~c_2$ and $\lambda$ solely.

\begin{figure}[!h]
\subfigure{
\begin{minipage}[t]{0.5\linewidth}
\centering
\includegraphics[height=.23\textheight,width=.34\textheight, angle =0]{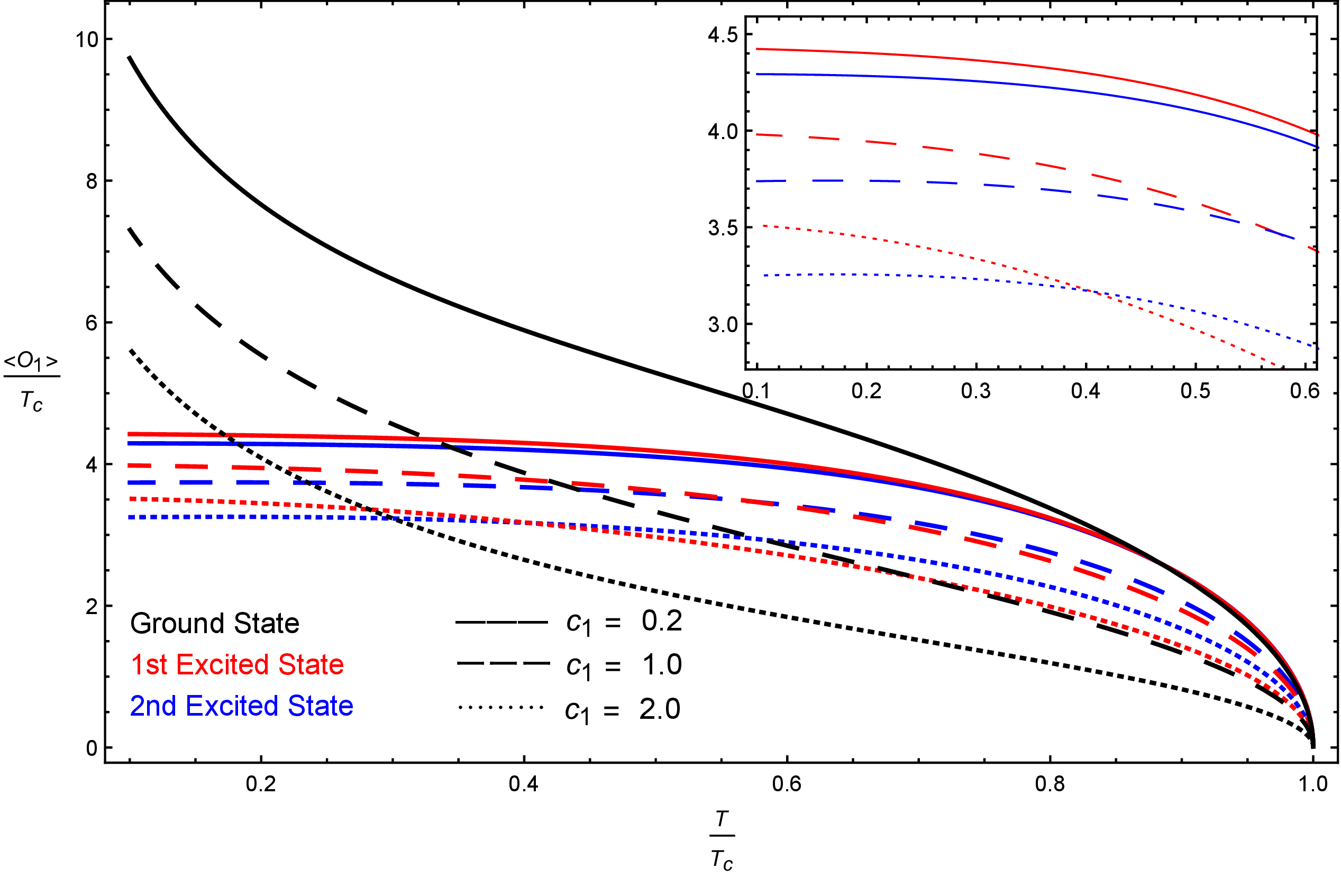}
\includegraphics[height=.23\textheight,width=.34\textheight, angle =0]{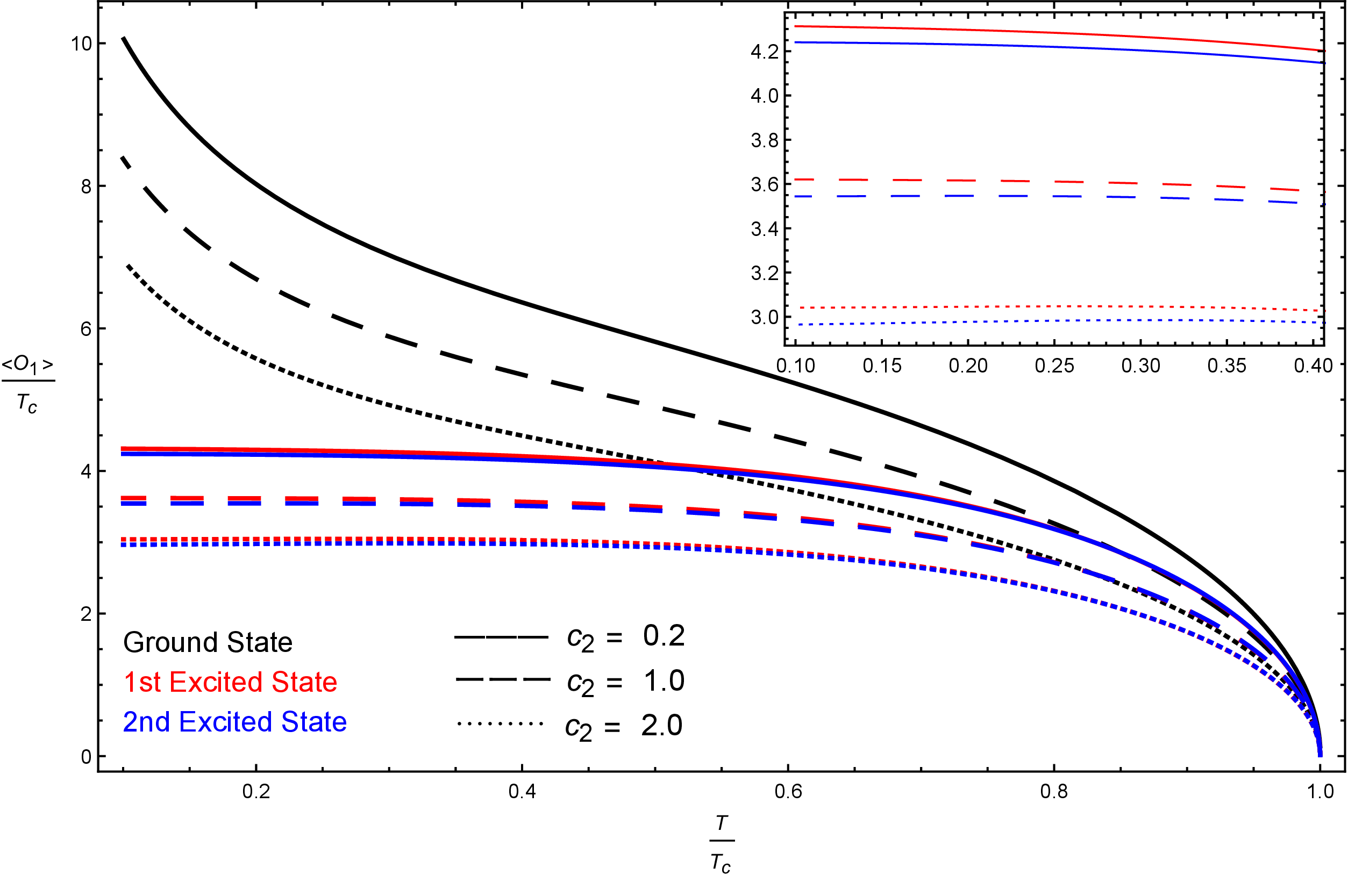}
\includegraphics[height=.23\textheight,width=.34\textheight, angle =0]{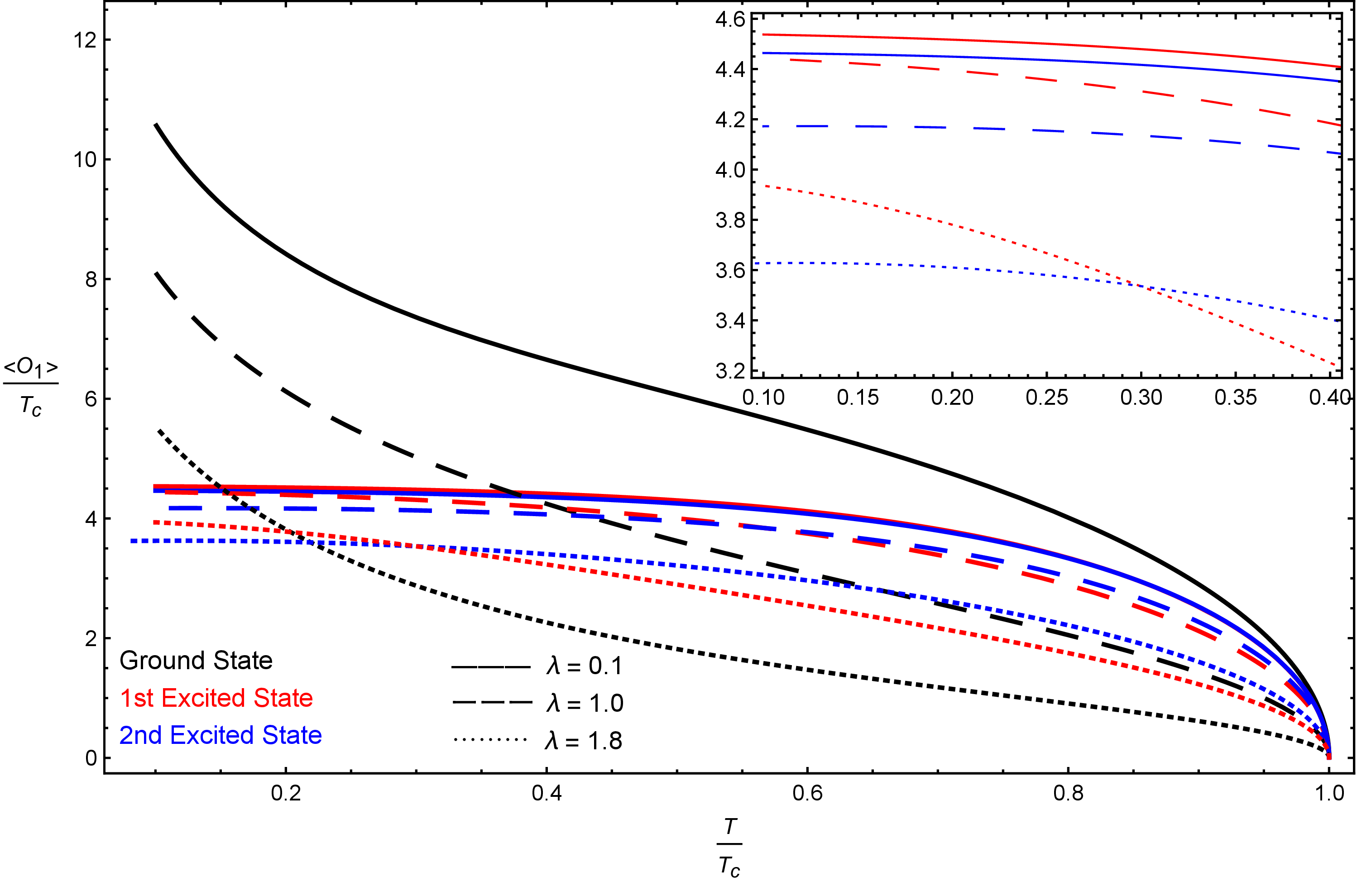}
\end{minipage}
}
\subfigure{
\begin{minipage}[t]{0.5\linewidth}
\centering
\includegraphics[height=.23\textheight,width=.34\textheight, angle =0]{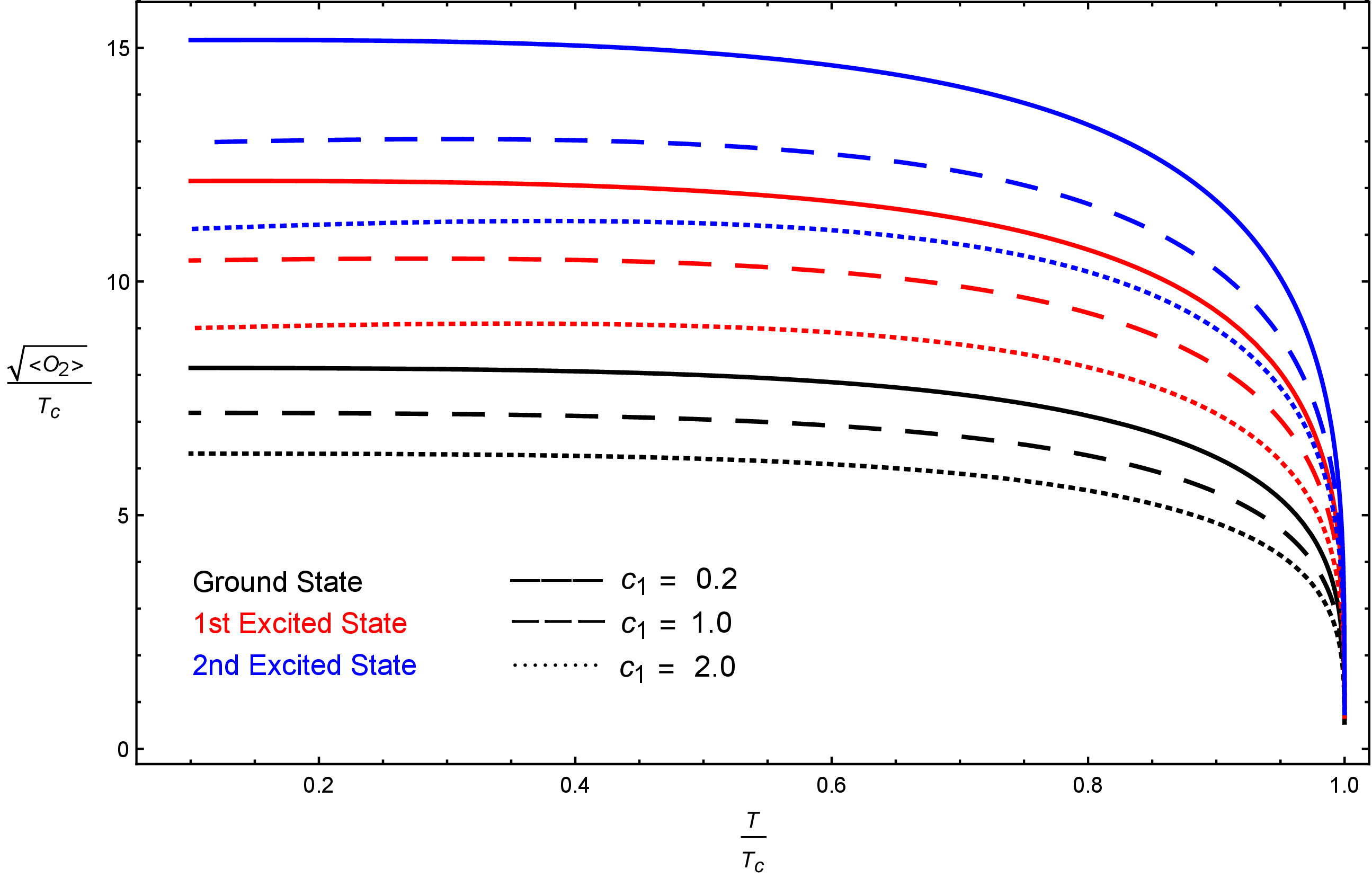}
\includegraphics[height=.23\textheight,width=.34\textheight, angle =0]{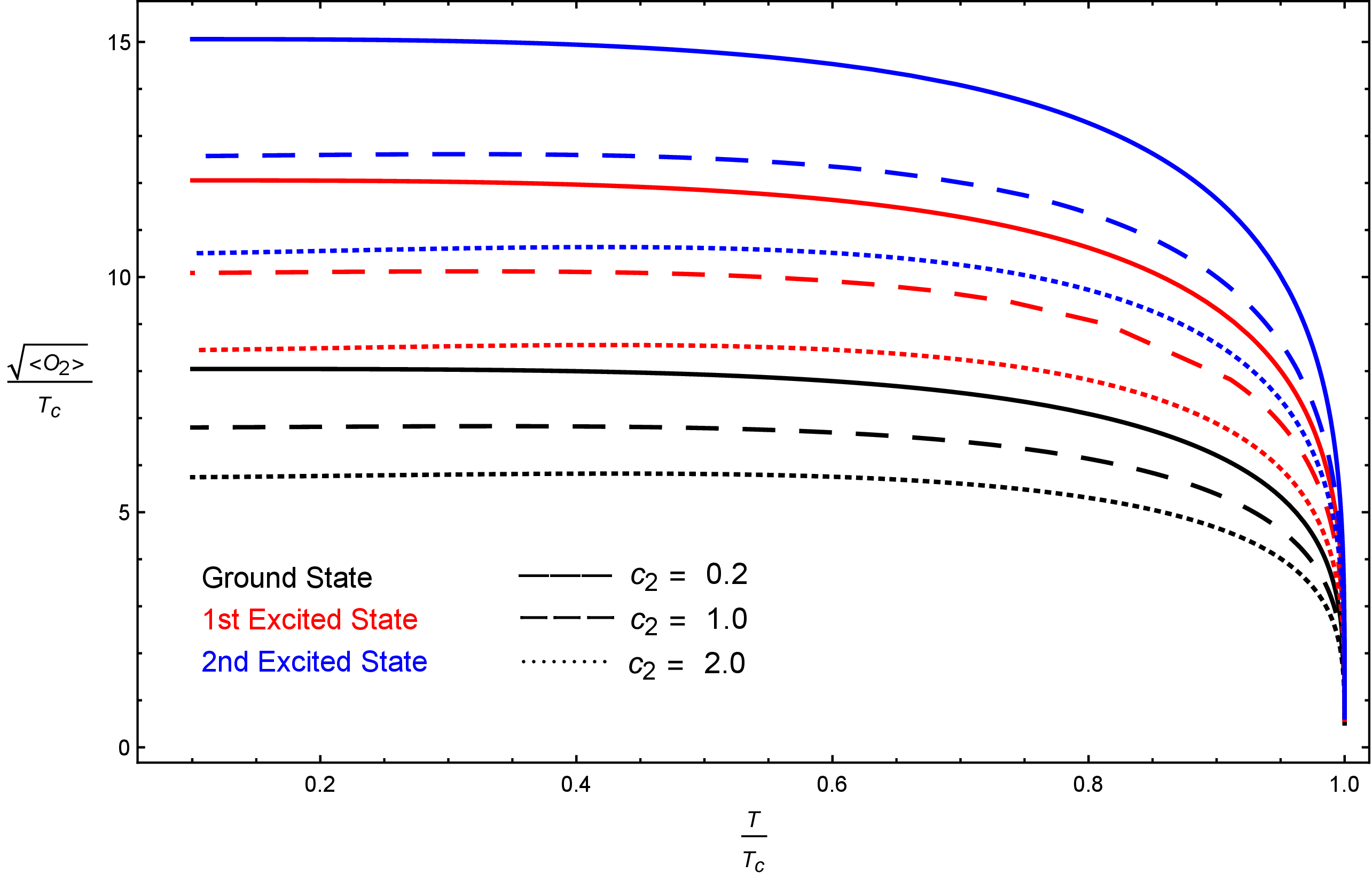}
\includegraphics[height=.23\textheight,width=.34\textheight, angle =0]{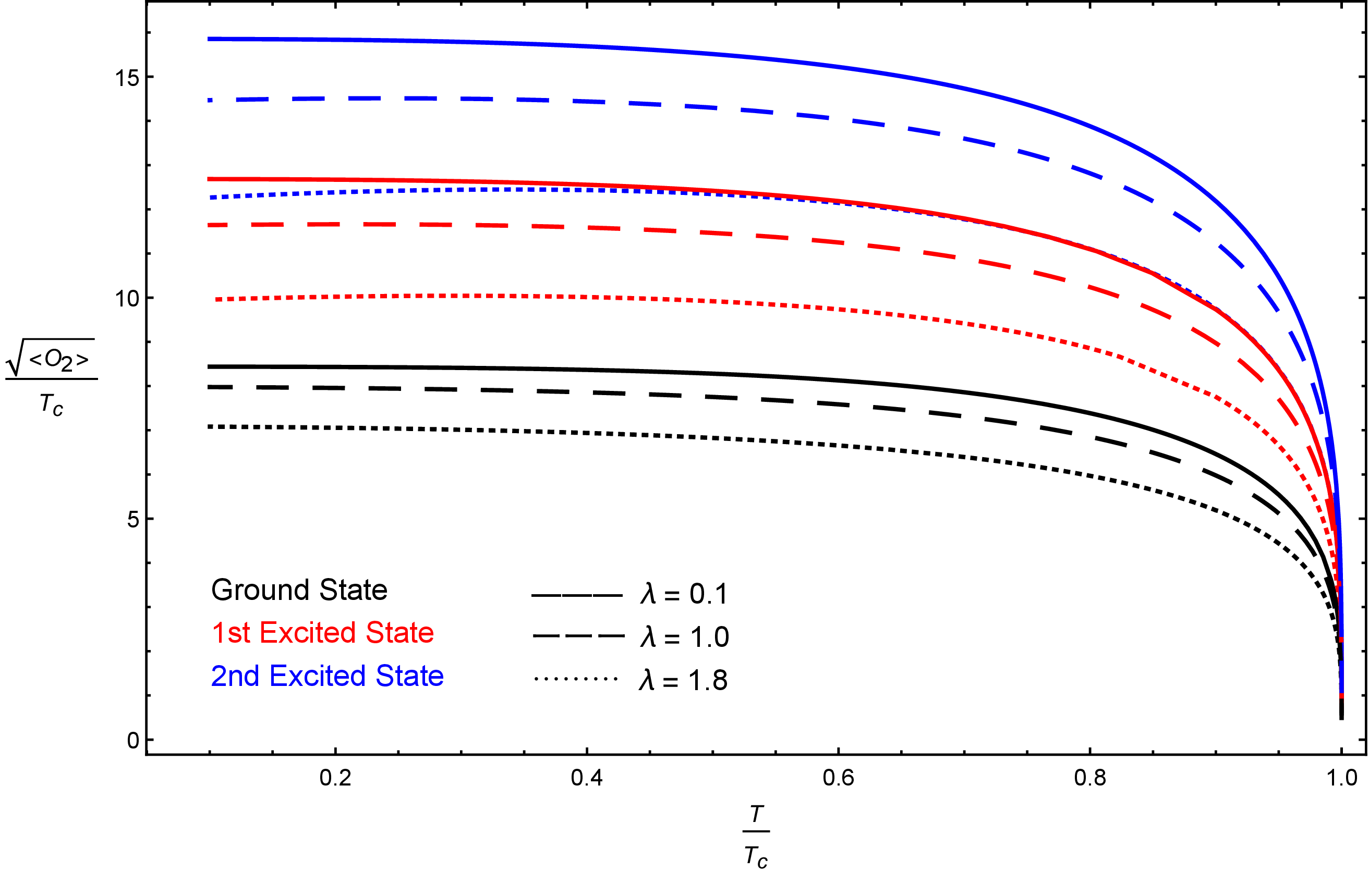}
\end{minipage}
}

\centering
\caption{$\langle \ocal_1 \rangle$ and $\langle \ocal_2 \rangle$ condensates in various configurations of massive gravity where the $\langle \ocal_1 \rangle$ and $\langle \ocal_2 \rangle$ condensates are presented in left and right panel, respectively. From top to bottom the results correspond to changing parameters $c_1,~c_2$ and $\lambda$ accordingly, where the condensations of ground states, first and second excited states are denoted by black lines, red and blue lines. In all plots, the parameters from large to small sequencing is marked by dotted lines, dashed lines and solid lines. } \label{o 1 2con}
\end{figure}

In Fig. \ref{o 1 2con}, we present our condensation results in various configurations of massive gravity background, where the $\langle \ocal_1 \rangle$ and $\langle \ocal_2 \rangle$ condensates are presented in the left and the right panel, respectively. In the first row we fix $c_2=0,~\lambda=1$ and study the coupling coefficient $c_1$ in the cases of $c_1=0.2,~c_1=1.0,~c_1=2.0$. In the middle row, we also study the coupling coefficient $c_2$ in the same way while setting $c_1=0,~\lambda=1$. In the last row, we fix $c_1=1,~c_2=-0.5$ while tuning the graviton mass from $\lambda=0.1$ to $\lambda=1.8$.

From the three plots of $\langle \ocal_1 \rangle$ condensates, we can see that the condensates of ground states rise rapidly at $T/T_c=1$ and then go to infinity at lower temperatures. Such singularity indicates that the charge of the scalar field is not large enough at low temperatures and the consideration of the backreaction is needed. The excited states also begin to condense around $T_c$, however, with the condensation values converge to limited values at low temperatures. We can also see that the condensation values of the ground states are larger than the excited states when the coupling coefficients $c_1,~c_2$ and the graviton mass $\lambda$ are small. Since, generally, the critical temperatures of the ground states are much higher than the excited states, such phenomenon indicates that the superconducting state can be reached firstly through the ground states, and then, the excited states, which is in accord with various experiments. However, when the effect of massive gravity is stronger, the condensation values of the ground states can be equal or even less than the excited states. Moreover, from the sub-figures shown inside the first and third plots in the left panel, one can find that, with larger coupling coefficients and graviton mass, the condensation values of the second excited states are higher than the first excited states when the temperature is high. At $T\rightarrow 0$, the scalar condensates of the second excited states converge to lower values comparing to the first excited states. In addition, the condensation values are obviously lowered as we enforce the effects of massive gravity, therefore, these quantities can not be set too large, or the superconductivity would certainly not appear.

For the $\langle \ocal_2 \rangle$ condensates in the three plots of the right panel, the condensation values of all three plots eventually converge to constants at low temperatures, where the condensation values of higher states are larger than their former states. Besides, it is predictable that the condensates would disappear if one continues to enlarge the coupling coefficients and the graviton mass.

It is obvious that our model also possesses a second order transition from normal state to superconducting state near the critical point, which is a square root behaviour predicted by the mean field theory. Therefore, we fit the condensation curves $vs.$ temperatures for both $\ocal_1$ and $\ocal_2$ operators near the critical temperatures according to our results shown in Fig. \ref{o 1 2con}.

\begin{figure}[!t]
\centering
\includegraphics[height=.15\textheight,width=.21\textheight, angle =0]{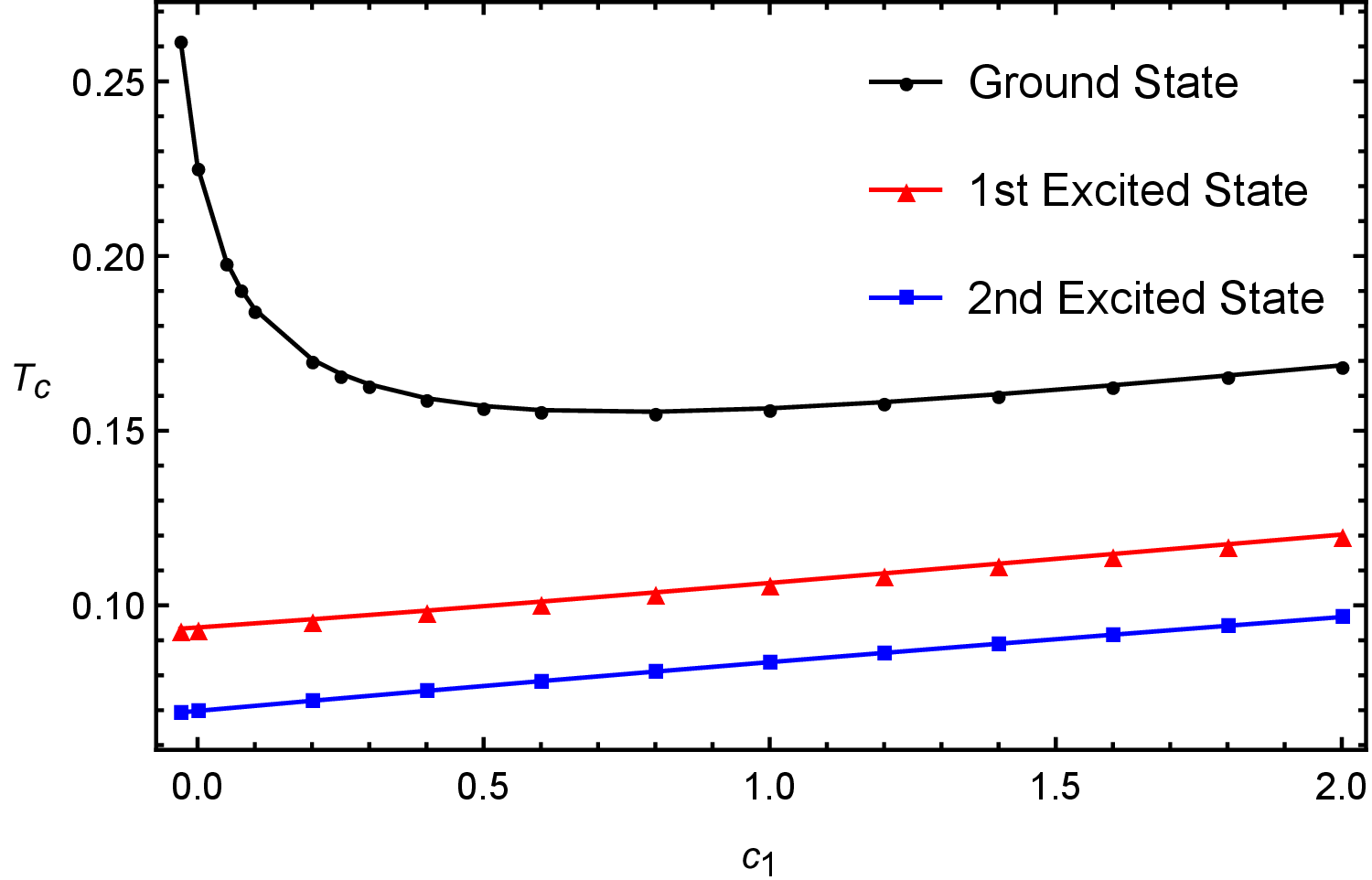}
\includegraphics[height=.15\textheight,width=.21\textheight, angle =0]{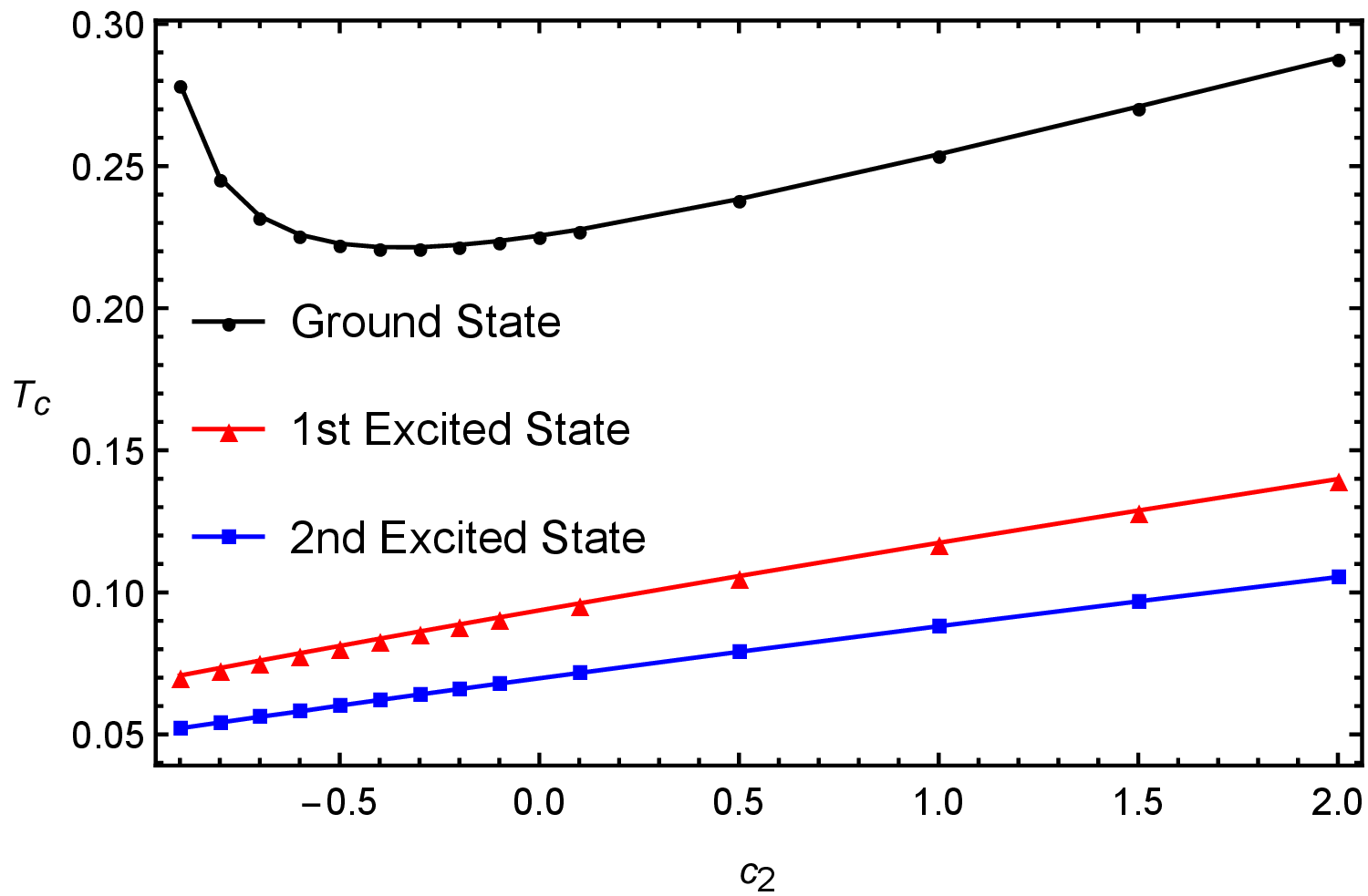}
\includegraphics[height=.15\textheight,width=.21\textheight, angle =0]{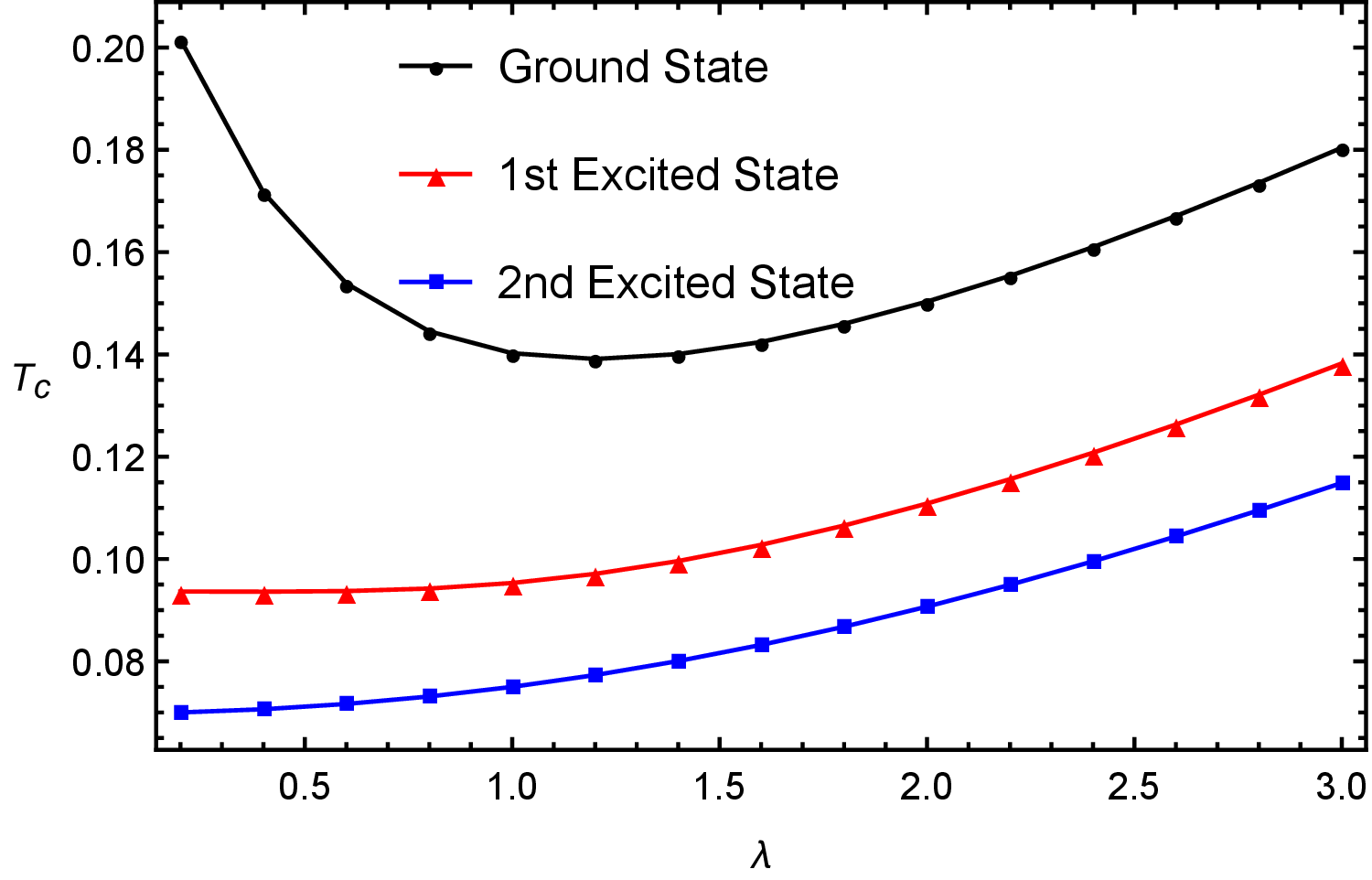}\\
\includegraphics[height=.15\textheight,width=.21\textheight, angle =0]{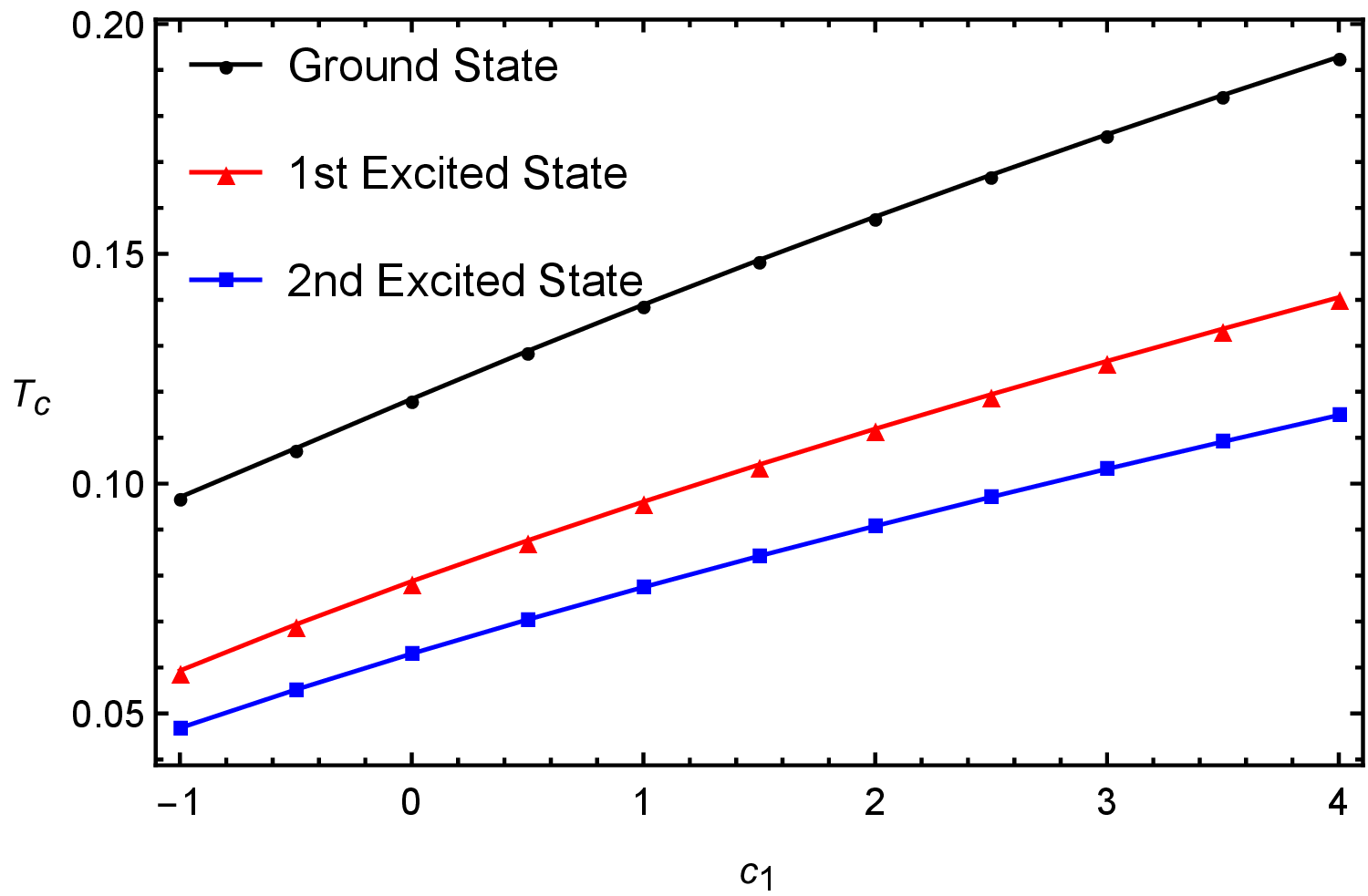}
\includegraphics[height=.15\textheight,width=.21\textheight, angle =0]{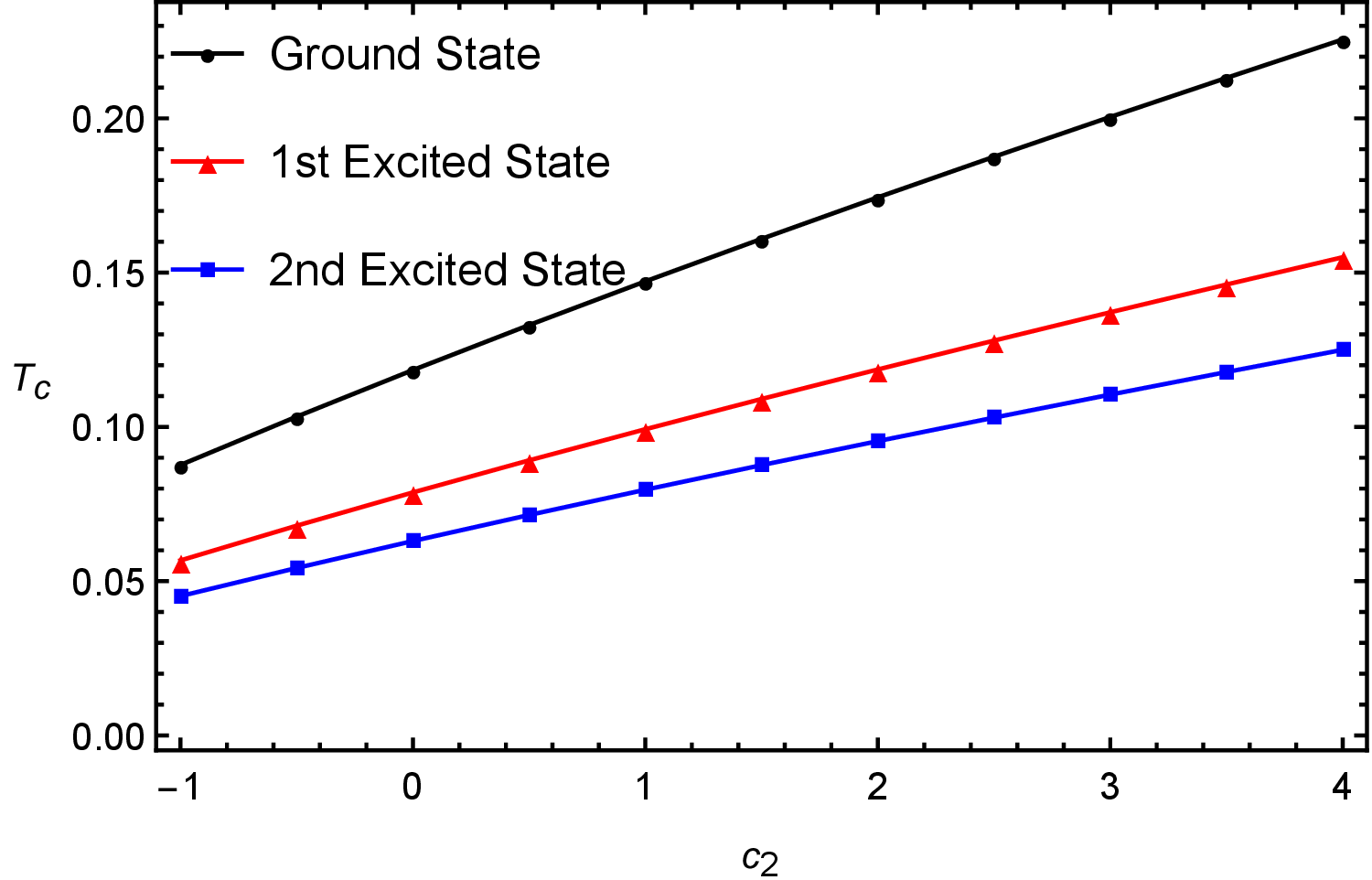}
\includegraphics[height=.15\textheight,width=.21\textheight, angle =0]{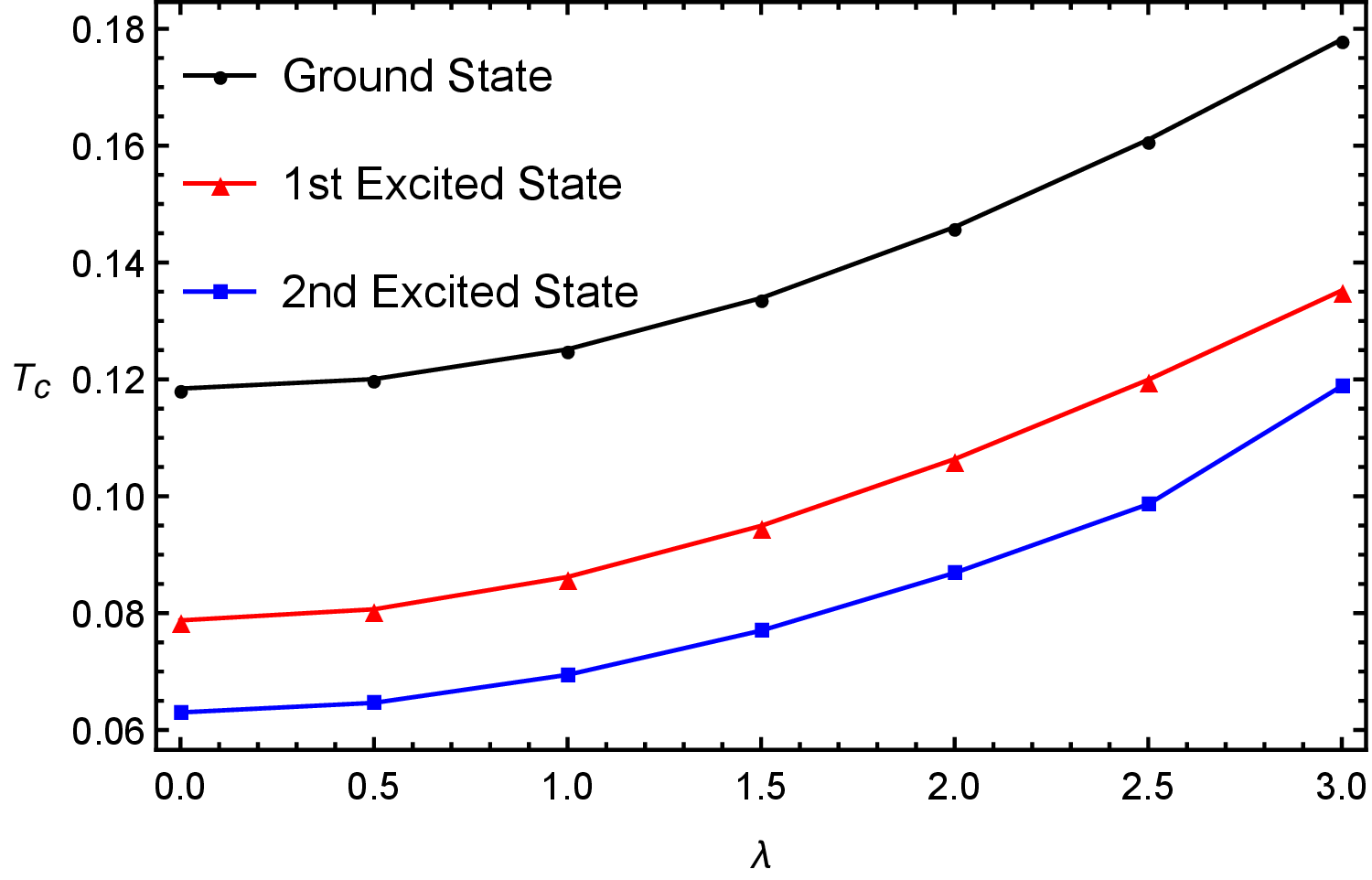}
\caption{Numerical results of critical temperatures for both kinds of condensates, where the three plots in the first row indicate the results of $ \ocal_1 $ operator and the second row correspond to the $ \ocal_2 $ operator. In all plots, the black, red and blue lines correspond to the critical temperatures of ground states, first and second excited states, respectively.}\label{Tc}
\end{figure}

The formats of the fittings are as follow:
\begin{equation}
\langle \ocal_1 \rangle \approx \zeta^{(n)}~ T_{c}^{(n)}(1-T/T_{c}^{(n)})^{1/2},
\end{equation}
\begin{equation}
\langle \ocal_2 \rangle \approx \zeta^{(n)}~ (T_{c}^{(n)})^{2}(1-T/T_{c}^{(n)})^{1/2},
\end{equation}
where $\zeta^{(n)}$ and $T_{c}^{(n)}$ are the fitting coefficient and critical temperature of $n$-th state. The results of $\zeta^{(n)}$ are presented in Tab. \ref{table zeta}. The corresponding critical temperatures are shown at Fig. \ref{Tc} where we have also calculated other critical temperatures in different massive gravity backgrounds to show their tendencies.
\begin{table}[!h]
\begin{center}
\begin{tabular}{|c| c c c| c c c | c c c |c c}
\hline
\multirow{2}*{$Case$} &\multirow{2}*{$c_1$} & \multirow{2}*{$c_2$} & \multirow{2}*{$\lambda$} & \multicolumn{3}{c|}{$\langle \ocal_1 \rangle$} & \multicolumn{3}{|c|}{$\langle \ocal_2 \rangle$}\\
\cline{5-10}
& & & & $\zeta^{(0)}$ & $\zeta^{(1)}$ & $\zeta^{(2)}$ & $\zeta^{(0)}$ & $\zeta^{(1)}$ & $\zeta^{(2)}$ \\
\hline
$\romannumeral1$ &0.2 & 0 & 1 & 7.71 & 7.87 & 7.97 & 134 & 314 & 497 \\
\cline{5-10}
$\romannumeral2$&1.0 & 0 & 1 & 4.21 & 6.21 & 6.69 & 111 & 240 & 368 \\
\cline{5-10}
$\romannumeral3$&2.0 & 0 & 1 & 2.58 & 4.55 & 5.37 & 82 & 182 & 293 \\
\hline
$\romannumeral4$&0 & 0.2 & 1 & 8.92 & 7.84 & 7.90 & 138 & 305 & 486 \\
\cline{5-10}
$\romannumeral5$&0 & 1.0 & 1 & 7.53 & 6.61 & 6.67 & 101 & 236 & 351 \\
\cline{5-10}
$\romannumeral6$&0 & 2.0 & 1 & 6.34 & 5.66 & 5.67 & 79 & 170 & 260 \\
\hline
$\romannumeral7$&1 & -0.5 & 0.1 & 9.23 & 8.01 & 8.03 & 146 & 333 & 527 \\
\cline{5-10}
$\romannumeral8$&1 & -0.5 & 1.0 & 4.51 & 6.82 & 7.37 & 123 & 285 & 457 \\
\cline{5-10}
$\romannumeral9$&1 & -0.5 & 1.8 & 1.92 & 3.89 & 5.11 & 98 & 209 & 330 \\
\hline
\end{tabular}\caption{Fitting coefficients $\zeta^{(n)}$ for both $\langle \ocal_1 \rangle$ and $\langle \ocal_2 \rangle$ condensation curves.}\label{table zeta}
\end{center}
\end{table}

From the fitting coefficients of $\langle \ocal_2 \rangle$ condensate listed in Tab. \ref{table zeta}, one can easily find that, similar to the models in Einstein gravity, each $\zeta^{(n)}$ of $n$-th state is higher than its former state in every case. And, as we enlarge the coupling coefficients or the graviton mass, $\zeta^{(n)}$ of each state decreases rapidly for both $\ocal_1$ and $\ocal_2$ operators. The difference between our model in the massive gravity framework and the model in Einstein gravity is reflected by the $\langle \ocal_1 \rangle$ condensate, where in the latter model the fitting coefficient $\zeta^{(n)}$ of $n$-th state is smaller than its former state around the critical temperature. Whereas, on the contrary, in our model around the critical temperature, the values of the scalar field condense larger in the $n$-th excited state than its former state as shown in the cases $\romannumeral1$ to $\romannumeral3$, the cases $\romannumeral8$ and $\romannumeral9$. One can also directly discover such tendency by looking at the first and the third plot around $T_c$ in the left panel of Fig. \ref{o 1 2con}. As for the four remaining cases from $\romannumeral4$ to $\romannumeral7$, we can see that the $\zeta$ of excited states are smaller than the ground states with $\zeta^{(2)}$ being slightly larger than $\zeta^{(1)}$. In addition, as we extend our calculations to forth excited state for the cases $\romannumeral4$ to $\romannumeral7$, $\zeta^{(n)}$ of the excited state increases slightly with $n$-th excited state.

In Fig. \ref{Tc}, we present our numerical results of critical temperatures for both $\ocal_1$ (first row) and $\ocal_2$ (second row) operators, where from left to right they correspond to the cases of tuning $c_1$ with fixing $c_2=0,~\lambda=1$, tuning $c_2$ with fixing $c_1=0,~\lambda=1$ and tuning $\lambda$ with fixing $c_1=1,~c_2=-0.5$, respectively. From the three plots in the first row, we can indeed see that the tendencies of $T_c$ of ground states are from high values to minimums and then to high values again, as the coupling coefficients and the graviton mass are tuned from small to large. It is noteworthy that, for both $\langle \ocal_1 \rangle$ and $\langle \ocal_2 \rangle$ condensates where the effect of massive gravity is strong, the holographic superconductor possesses higher critical temperatures than the model in general relativity. However, this does not indicate that the $T_c$ can be infinitely high by simply enlarging these parameters since, as we can also see from Fig. \ref{o 1 2con}, the condensation values become small when enforcing the couplings or the graviton mass itself. Meanwhile, by lowering the coupling coefficients to minus values, $T_c$ can not be infinitely high as well. Since minus $c_1$ and $c_2$ who determine the temperature of the black hole, with large absolute values would lead to minus black hole temperature. In addition, for the $\langle \ocal_1 \rangle$ condensate of the ground state shown at Fig. \ref{o 1 2con} (left panel), the divergence at low temperature indicate that the charge of the scalar field is not big enough such that the backreaction on the metric will have to took into consideration.
\subsection{Conductivity}
In this subsection we study  the effects of massive gravity on the conductivity in the ground and the excited states of our holographic superconductor.

By turning on perturbations of the vector potential $A_x$ in the bulk geometry of Schwarzschild-AdS black hole, we consider the Maxwell equation with a time dependence of $e^{- i \w t}$, the linearized equations are given as
\be\label{eq:Axeq}
A_x'' + \frac{f'}{f} A_x' + \left(\frac{\w^2}{f^2} - \frac{2
\psi^2}{f} \right) A_x = 0 \,.
\ee
When imposing the ingoing boundary conditions at the horizon, the asymptotic behaviour of the Maxwell field on the boundary is
\be
A_x = A_x^{(0)} + \frac{A_x^{(1)}}{r} + \cdots.
\ee
The conductivity can be obtained according to the Ohm's law
\be\label{eq:conductivity}
\sigma(\w) = - \frac{i A_x^{(1)}}{\w A_x^{(0)}} \,.
\ee
\begin{figure}[h]
\centering

\subfigure[]{
\begin{minipage}[t]{0.5\linewidth}
\centering
\includegraphics[height=.23\textheight,width=.34\textheight, angle =0]{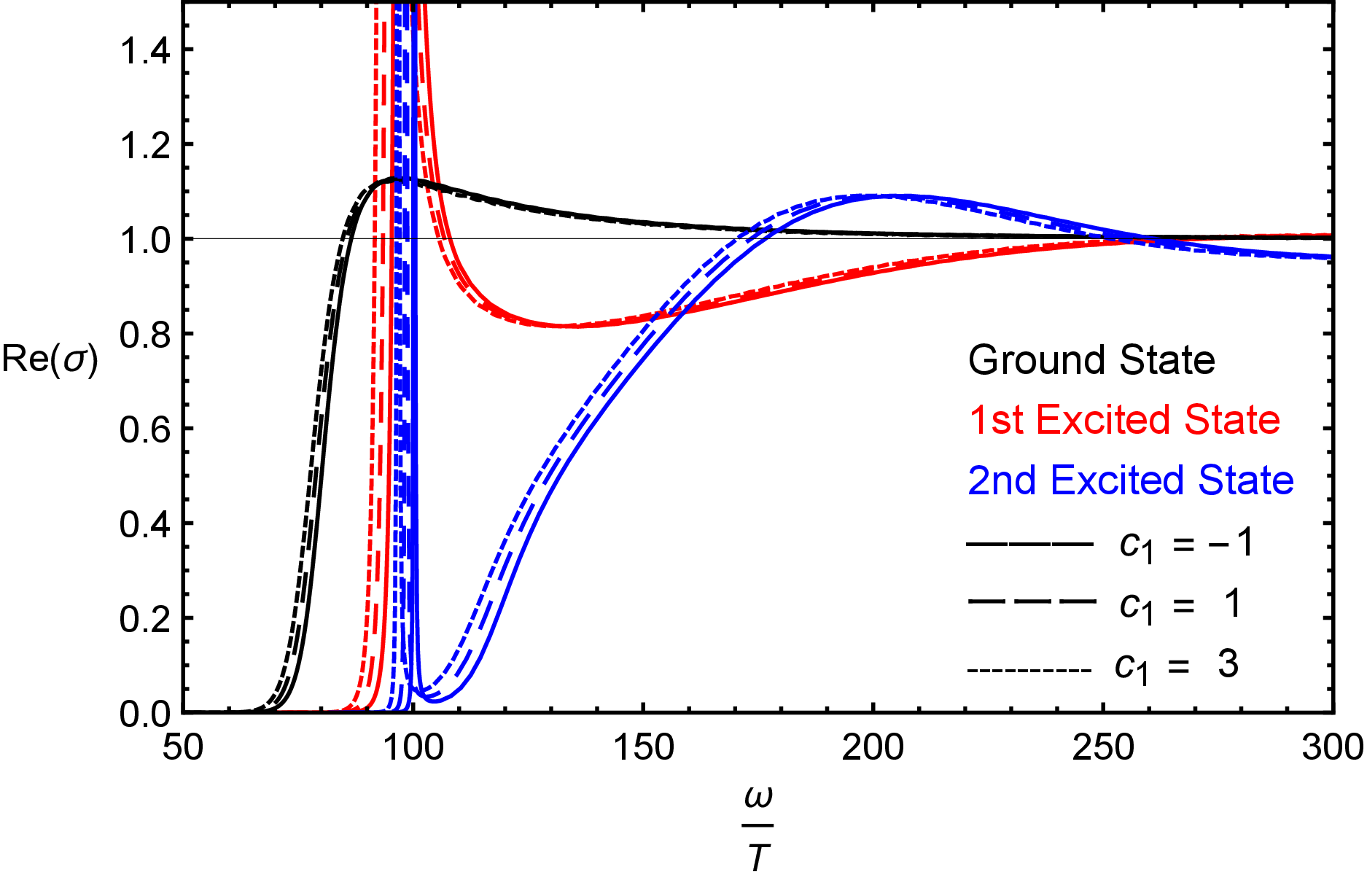}
\end{minipage}%
}%
\subfigure[]{
\begin{minipage}[t]{0.6\linewidth}
\centering
\includegraphics[height=.23\textheight,width=.34\textheight, angle =0]{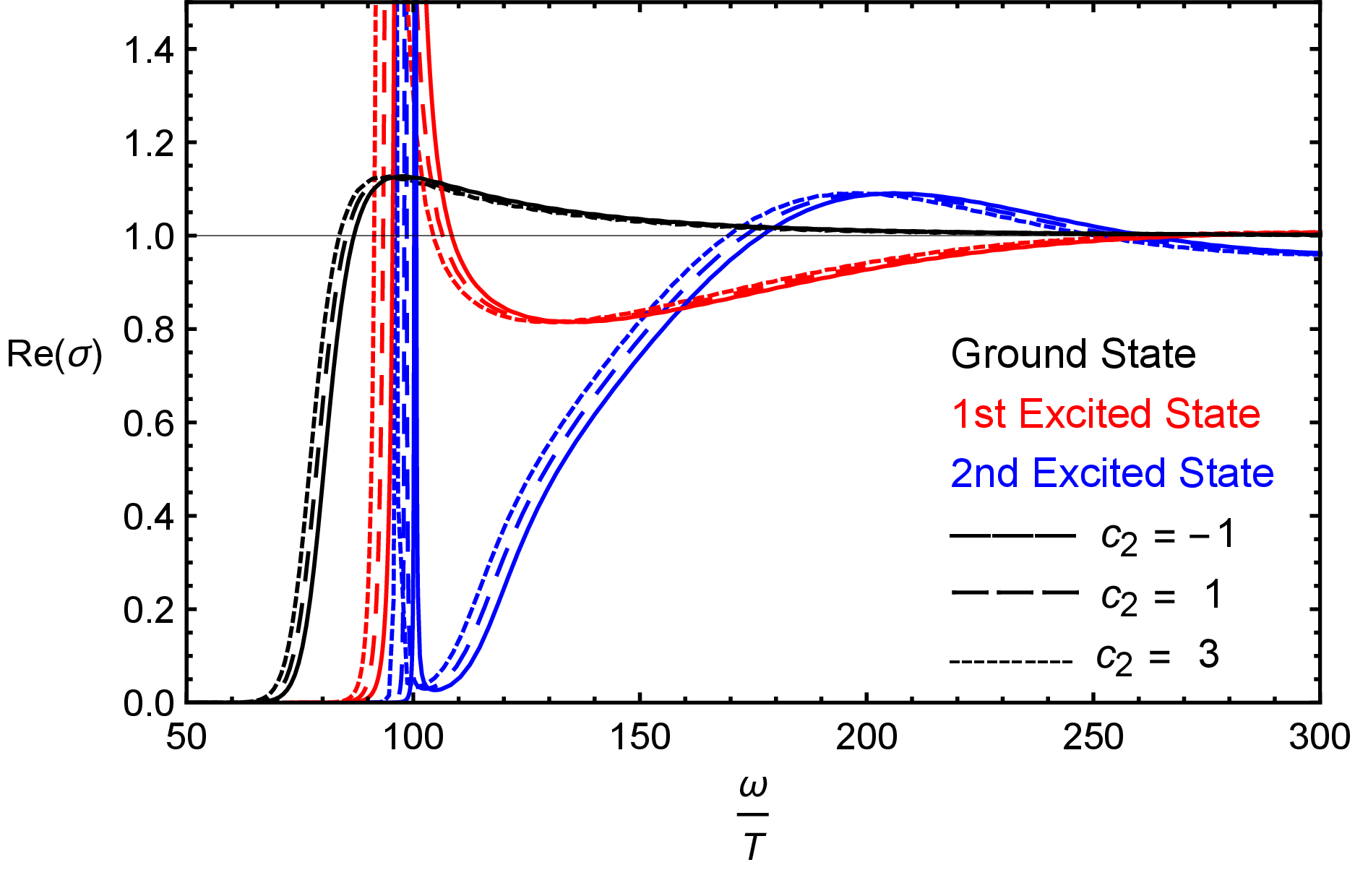}
\end{minipage}%
}%

\subfigure[]{
\begin{minipage}[t]{0.65\linewidth}
\centering
\includegraphics[height=.23\textheight,width=.34\textheight, angle =0]{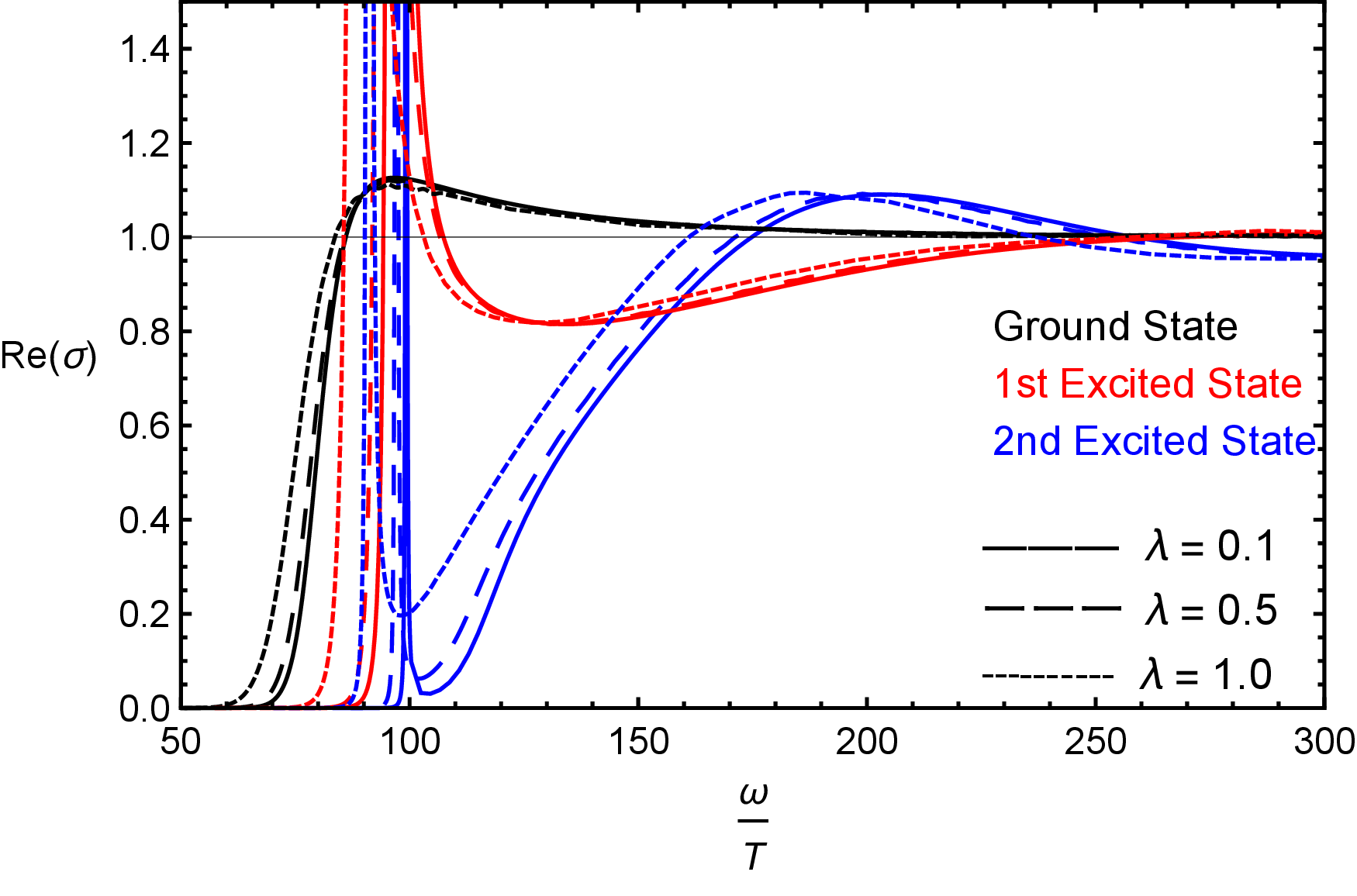}
\end{minipage}
}%

\centering
\caption{The real part of optical conductivity, where (a), (b) and (c) correspond to the studies of $c_1$, $c_2$ and $m$ respectively. The black, red and blue lines represent ground states, first and second excited states. In all three plots, the studied parameters from large to small sequencing is marked by short dashed lines, long dashed lines and solid lines. }\label{axr}
\end{figure}

\begin{figure}[!h]
\centering

\subfigure[]{
\begin{minipage}[t]{0.5\linewidth}
\centering
\includegraphics[height=.23\textheight,width=.34\textheight, angle =0]{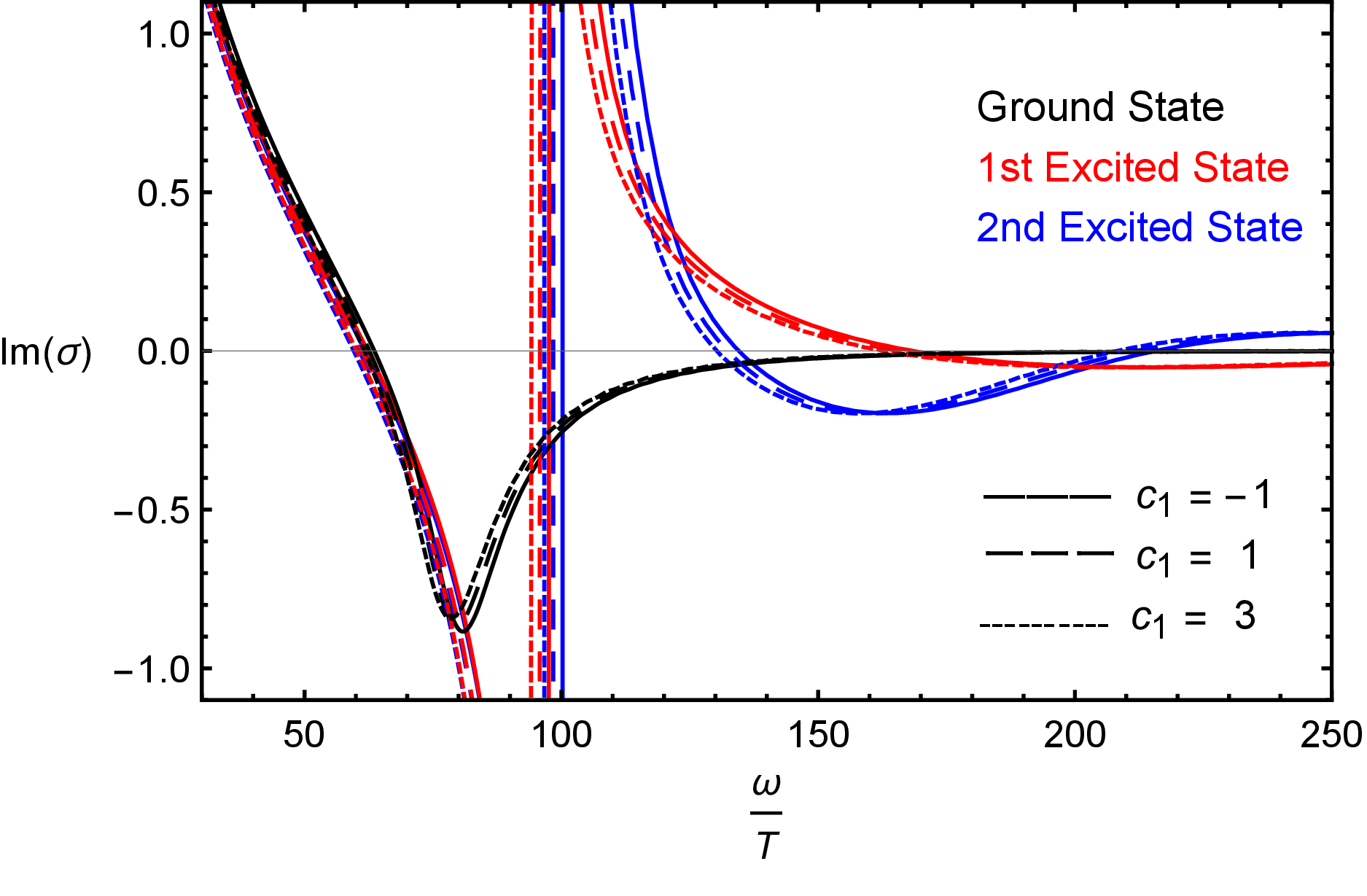}
\end{minipage}%
}%
\subfigure[]{
\begin{minipage}[t]{0.6\linewidth}
\centering
\includegraphics[height=.23\textheight,width=.34\textheight, angle =0]{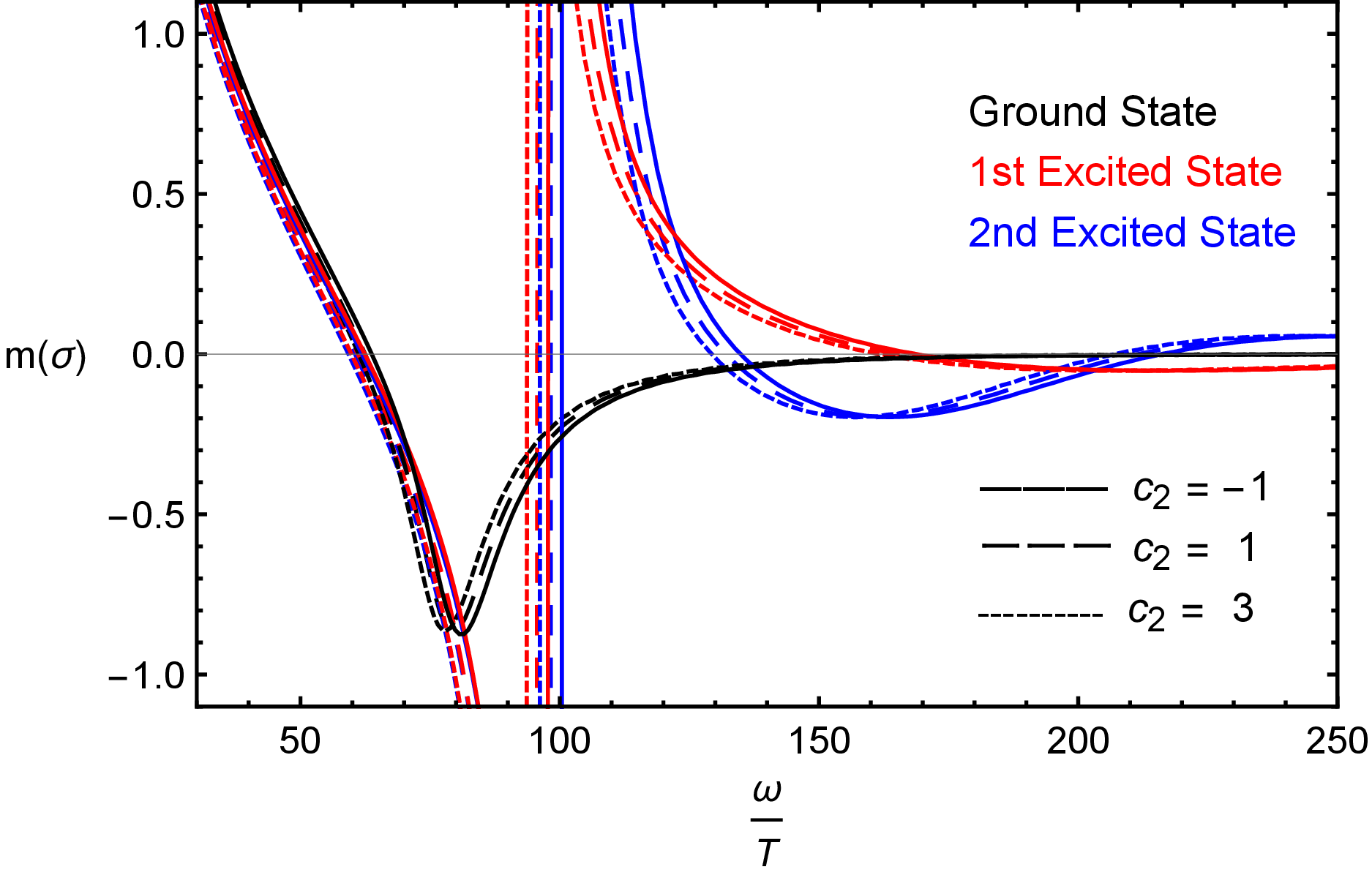}
\end{minipage}%
}%

\subfigure[]{
\begin{minipage}[t]{0.65\linewidth}
\centering
\includegraphics[height=.23\textheight,width=.34\textheight, angle =0]{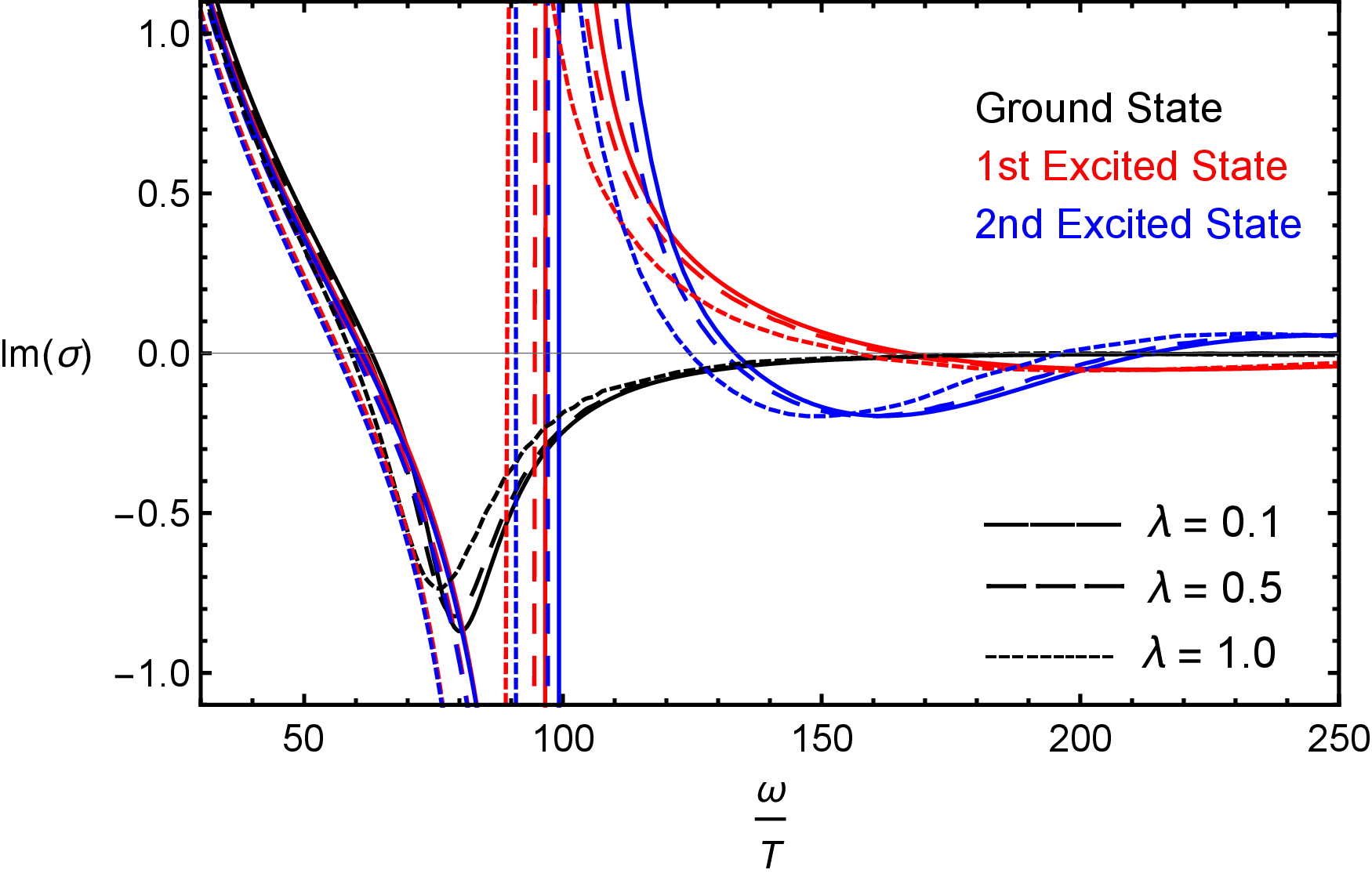}
\end{minipage}
}%

\centering
\caption{The imaginary part of optical conductivity, where (a), (b) and (c) correspond to the studies of $c_1$, $c_2$ and $\lambda$ respectively. The black, red and blue lines represent ground states, first and second excited states. In all three plots, the studied parameters from large to small sequencing is marked by short dashed lines, long dashed lines and solid lines. }\label{axi}
\end{figure}

In Fig. \ref{axr} and Fig. \ref{axi}, we plot the real and imaginary part of optical conductivity as a function of frequency at low temperature $T/T_c \approx 0.100$ for operator $\ocal_2 $ of ground states and excited states. In both figures, we use black lines, red lines and blue lines to represent the ground states, first and second excited states. The studied parameters from large to small sequencing is denoted by short dashed lines, long dashed lines and solid lines. In both Fig. \ref{axr} and Fig. \ref{axi}, plots. (a) and (b), we study the cases $c_1=-1,~c_1=1,~c_1=3$ with $c_2$ and $\lambda$ fixed at $c_2=0,~\lambda=0.2$ and the cases $c_2=-1,~c_2=1~c_2=3$ with $c_1$ and $\lambda$ fixed at $c_1=0,~m=0.2$ respectively. In plot. (c), we study the cases $\lambda=0.1,~\lambda=0.3,~\lambda=0.5$ with $c_1$ and $c_2$ fixed at $c_1=1,~c_2=-0.5$.

As it is known to all that in the real, dissipative part of conductivity, the horizontal lines at $\text{Re}[\sigma]=1$ correspond to temperatures higher than the critical temperature $T_c$ where there is no scalar condensate.
In the three plots of Fig. \ref{axr}, the gaps of ground states (black lines) open at $\omega/T\approx 60$ till the $\text{Re}[\sigma]$ exponentially converge to $1$.
Since we study the optical conductivity at $T/T_c\rightarrow 0$, the peaks of the ground states develop to almost delta functions as they are generated to excited states.
Besides, comparing with the $\text{Re}[\sigma]$ in the ground states, there exist one additional peak for the first excited state and two additional peaks for the second excited state.
Moreover, we find that the number of additional peaks of $n$-th excited state is equal to $n$. Similar behaviours can also be found at the imaginary part of the conductivity as shown at Fig. \ref{axi}.

In both figures, we can see that these studied parameters only have slight effects on the optical conductivity. By enlarging the three parameters, each state of both the $\text{Re}[\sigma]$ and $\text{Im}[\sigma]$ slightly shift to lower frequencies.

\subsection{Conclusions and discussions}\label{conclusion}
In this paper, we investigated a holographic superconductor model in the dRGT nonlinear gravity  which is constructed by Maxwell field coupled to a massive scalar field in four-dimensional AdS spacetime. The effects of massive gravitons are involved in the black hole solution and are tuned by the three parameters $c_1, c_2$ and $\lambda$. We noticed that the effects of massive graviton would not appear and the solutions would go back to \cite{Hartnoll:2008vx} in some situations. We managed to study the effects of $c_1,~c_2$ and $\lambda$ in situations that $c_1$ and $c_2$ can not be zero simultaneously and $\lambda \neq 0$. With two of them remained fixed, we varied one of these parameters for each time to study their effects on condensate and conductivity separately.
Our results show that, similar to the holographic superconductor in the general relativity, the critical chemical potential $\mu_c$ in higher excited states also possess larger values where the difference between consecutive states is also about five for both $\ocal_1$ and $\ocal_2$ operators.
Moreover, as we tune the coupling coefficients $c_1,~c_2$ and the graviton mass $\lambda$ from small to large, in the ground states, the critical temperatures $T_c$ of $\langle \ocal_1 \rangle$ decline from high values and then go upward. For the excited states, the $T_c$ increase with the growth of couplings factors and the graviton mass monotonously. For the $\ocal_2$ operator, the $T_c$ of each states possess higher values where the effect of gravity is stronger.
In addition, although there exists higher $T_c$ in our holographic superconductor, increasing $c_1,~c_2$ and $\lambda$ would dramatically lower the values of $\langle \ocal_1 \rangle$ and $\langle \ocal_2 \rangle$ condensates for each state.
As for the conductivity, the gaps of each state slightly shift to lower frequencies by enlarging the coupling coefficients and the graviton mass. Moreover, the profiles of both $\text{Re}[\sigma]$ and $\text{Im}[\sigma]$ show that the effects of the two coupling coefficients are almost the same. In the end, we find that there exists additional peaks for the conductivity of excited states, where the number of additional peaks is equal to $n$-th excited state.

There could be many interesting extensions of our work. First, as we have noticed in Fig. \ref{o 1 2con} that the large scalar field charge limit could not suffice for the ground states of $\langle \ocal_1 \rangle$ condensate, which required us to solve the coupled differential equations with Einstein equation. Second, around the critical temperature, the condensations of $\ocal_1$ operator perform differently comparing with the model in Einstein gravity and the mechanism is unclear. It is interestingly to answer this question with semi-analytical method \cite{Qiao:2020fiv}. In the end, we would like to extend our study of excited states in massive gravity to the p-wave and d-wave holographic superconductor in the future.

\subsection*{Acknowledgements}
We would like to thank Tong-Tong Hu for helpful discussion. Parts of computations were performed on the   shared memory system at  institute of computational physics and complex systems in Lanzhou university. This work was supported by the Fundamental Research Fund for Physics of Lanzhou University (No. Lzujbky-2019-ct06).

\end{document}